%% file: main.tex
\documentclass[acmsmall,screen]{acmart}
\input{config/package}
\input{config/definedcmd}
\begin{document}
\input{config/topmatters}
\begin{abstract}
    \input{src/00-abstract}

\end{abstract}
\maketitle
\input{src/notations}
\input{src/01-introduction}

\input{src/02-preliminary}
\input{src/03-approach}
\input{src/04-evaluation}
\input{src/05-related}

\input{src/06-conclusion}
\input{src/10-statement}
\input{src/10-acknowledgement}

\bibliographystyle{ACM-Reference-Format}
\bibliography{src/reference}

\end{document}

%% file: config/package.tex
\usepackage{tcolorbox}
\usepackage{xspace}
\usepackage{enumitem}
\usepackage{caption}
\usepackage{subcaption}
\usepackage{bbding}
\usepackage{url}
\usepackage{booktabs}
\usepackage{tabularx}
\usepackage{circledsteps} 

\usepackage{multicol}

\usepackage{makecell}
\usepackage{multirow}
\newcolumntype{C}[1]{>{\centering\arraybackslash}p{#1}}

\usepackage{amsmath,bm,amsthm}
\theoremstyle{definition}
\newtheorem{exmp}{Example}[section] 

\usepackage{algorithm}
\usepackage{algpseudocode}
\algnewcommand\algorithmicforeach{\textbf{for each}}
\algdef{SE}[FUNCTION]{Function}{EndFunction}[2]{\algorithmicfunction\ \textnormal{#1}\ (#2)}{\algorithmicend\ \algorithmicfunction}%
\algdef{S}[FOR]{ForEach}[1]{\algorithmicforeach\ #1\ \algorithmicdo}
\algrenewcommand\alglinenumber[1]{\footnotesize #1}
\algrenewcommand\algorithmicrequire{\small \textbf{input:}}
\algrenewcommand\algorithmicensure{\small \textbf{output:}}
\algrenewcommand\algorithmicfunction{\textbf{Function}}
\algtext*{EndFunction} %
\algtext*{EndIf} %
\algtext*{EndFor} %
\algtext*{EndWhile} %

\usepackage{fontawesome} 
\usepackage{xcolor}
\usepackage{framed}
\usepackage{textcomp}

\usepackage{microtype}
\usepackage{ifthen}

%% file: config/definedcmd.tex
\newcounter{rq}
\newenvironment{answertorq}{%
    \begin{tcolorbox}[
            arc=2mm,
            boxrule=0.5pt,
            left=2pt,
            right=2pt,
            top=2pt,
            bottom=2pt
        ]
    \textbf{Answer to RQ\refstepcounter{rq}\therq{}:}
}{\end{tcolorbox}}

\tcbuselibrary{breakable}

\ifthenelse{\isundefined{\editmode}}{
    \newcommand{\editnote}[3]{%
    }
}
{
    \newcommand{\editnote}[3]{\xspace\colorbox{#1}{\sffamily \smaller \textcolor{white}{~\faCommenting{}~#2~}}\textcolor{#1}{~#3}\xspace}
}
\definecolor{latte-teal}{RGB}{23, 146, 153}
\definecolor{nord0}{HTML}{2E3440}
\definecolor{nord1}{HTML}{3B4252}
\definecolor{nord2}{HTML}{434C5E}
\definecolor{nord3}{HTML}{4C566A}
\definecolor{nord4}{HTML}{D8DEE9}
\definecolor{nord5}{HTML}{E5E9F0}
\definecolor{nord6}{HTML}{ECEFF4}
\definecolor{nord7}{HTML}{8FBCBB}
\definecolor{nord8}{HTML}{88C0D0}
\definecolor{nord9}{HTML}{81A1C1}
\definecolor{nord10}{HTML}{5E81AC}
\definecolor{nord11}{HTML}{BF616A}
\definecolor{nord12}{HTML}{D08770}
\definecolor{nord13}{HTML}{EBCB8B}
\definecolor{nord14}{HTML}{A3BE8C}
\definecolor{nord15}{HTML}{B48EAD}

\newcommand{\congying}[1]{\editnote{olive}{Congying}{#1}}
\newcommand{\valerio}[1]{\editnote{red}{Valerio}{#1}}
\newcommand{\vale}[1]{\editnote{green}{Valerio}{#1}}
\newcommand{\scc}[1]{\editnote{violet}{S.C.}{#1}}
\newcommand{\jr}[1]{\editnote{brown}{jr}{#1}}
\newcommand{\hc}[1]{\editnote{latte-teal}{HC}{#1}}

\newcommand{\todo}[1]{\textcolor{black}{#1}}
\newcommand{\code}[1]{\texttt{#1}}
\newcommand{\tool}{\textsc{MR-Scout}\xspace}

\newcommand{\editedA}[1]{\textcolor{black}{#1}}
\newcommand{\editedT}[1]{\textcolor{black}{#1}}
\newcommand{\editedTS}[1]{\textcolor{black}{#1}}

%% file: config/topmatters.tex

\setcopyright{acmlicensed}
\acmJournal{TOSEM}
\acmYear{2024}
\acmVolume{1}
\acmNumber{1}
\acmArticle{1}
\acmMonth{3}
\acmPrice{15.00}
\acmDOI{---/---}

\renewcommand\footnotetextcopyrightpermission[1]{} 

\keywords{Software Testing, Metamorphic Testing, Metamorphic Relation, Automated Test Case Generation}

\ccsdesc[500]{Software and its engineering~Software testing and debugging}

\title{\tool: Automated Synthesis of Metamorphic Relations from Existing Test Cases}

\author{Congying Xu}
\orcid{0009-0000-2887-1690}
\email{cxubl@connect.ust.hk}
\affiliation{%
    \institution{The Hong Kong University of Science and Technology}
    \city{Hong Kong}
    \country{China}
}

\author{Valerio Terragni}
\orcid{0000-0001-5885-9297}
\email{v.terragni@auckland.ac.nz}
\affiliation{%
    \institution{The University of Auckland}
    \city{Auckland}
    \country{New Zealand}
}

\author{Hengcheng Zhu}
\orcid{0000-0002-3082-5957}
\email{hzhuaq@connect.ust.hk}
\affiliation{%
    \institution{The Hong Kong University of Science and Technology}
    \city{Hong Kong}
    \country{China}
}

\author{Jiarong Wu}
\orcid{0000-0001-6126-303X}
\email{jwubf@cse.ust.hk}
\affiliation{%
    \institution{The Hong Kong University of Science and Technology}
    \city{Hong Kong}
    \country{China}
}

\author{Shing-Chi Cheung}
\authornote{Shing-Chi Cheung is the corresponding author.}
\orcid{0000-0002-3508-7172}
\email{scc@cse.ust.hk}
\affiliation{%
    \institution{The Hong Kong University of Science and Technology}
    \city{Hong Kong}
    \country{China}
}

%% file: src/00-abstract.tex
\congying{[2023-08-17] ready to revise}
Metamorphic Testing (MT) alleviates the oracle problem by defining oracles based on metamorphic relations (MRs), that govern multiple related inputs and their outputs. 
However, designing MRs is challenging, as it requires domain-specific knowledge. This hinders the widespread adoption of MT. 
We observe that developer-written test cases can embed domain knowledge that encodes MRs. Such encoded MRs could be synthesized for testing not only their original programs but also other programs that share similar functionalities. 

In this paper, we propose \tool to automatically synthesize MRs from test cases in open-source software (OSS) projects. 
\tool first discovers MR-encoded test cases (MTCs), and then synthesizes the encoded MRs into parameterized methods (called \editedTS{\textit{codified MRs}}), and filters out MRs that demonstrate poor quality for new test case generation. 
\tool discovered over 11,000 MTCs from 701 OSS projects. 
Experimental results show that over \todo{97\%} of codified MRs are of high quality for automated test case generation, demonstrating the practical applicability of \tool. 
\editedT{
Furthermore, codified-MRs-based tests effectively enhance the test adequacy of programs with developer-written tests, leading to \todo{13.52\%} and \todo{9.42\%} increases in line coverage and mutation score, respectively.  
Our qualitative study shows that \todo{55.76\% to 76.92\%} of codified MRs are easily comprehensible for developers.
}

%% file: src/notations.tex
\newcommand{\tuple}[1]{\langle #1 \rangle}
\newcommand{\Rel}{\mathcal{R}}
\newcommand{\object}{ob}
\newcommand{\Output}{y}

\newcommand{\targetFunction}{f}
\newcommand{\targetFunctions}{F}
\newcommand{\targetFunctionsList}{\langle f_1, \cdots, f_m \rangle (m \geq1)}
\newcommand{\targetFunctionsSize}{|F|\geq1}
\newcommand{\programUnderTest}{P}
\newcommand{\classUnderTest}{CUT}
\newcommand{\detailedSourceAndFollowupInputs}{\langle x_1, \cdots, x_n \rangle}
\newcommand{\detailedSourceAndFollowupOutputs}{\langle f(x_1), \cdots,f(x_n)\rangle}
\newcommand{\detailedSourceInputs}{\langle x_1, \cdots, x_k\rangle}
\newcommand{\detailedFollowUpInputs}{\tuple{ x_{k+1}, \cdots, x_n}}
\newcommand{\detailedSourceOutputs}{\langle f(x_1), \cdots, f(x_k)\rangle}
\newcommand{\detailedFollowUpOutputs}{\tuple{ f(x_{k+1}), \cdots, f(x_n)}}

\newcommand{\shortedSourceInputs}{\underset{v=1\cdots k}{\langle x_v \rangle}}
\newcommand{\shortedFollowUpInputs}{\underset{w=(k+1)\cdots n}{\langle x_w \rangle}}
\newcommand{\shortedAllInputs}{\underset{i=1\cdots n}{\langle x_i \rangle}}
\newcommand{\shortedSourceOutputs}{\underset{v=1\cdots k}{\langle f(x_v) \rangle}}
\newcommand{\shortedAllOutputs}{\underset{i=1\cdots n}{\langle f(x_i) \rangle}}

\newcommand{\PshortedSourceOutputs}{\underset{v=1\cdots k}{\langle P(x_v) \rangle}}
\newcommand{\PshortedAllOutputs}{\underset{i=1\cdots n}{\langle P(x_i) \rangle}}

\newcommand{\wrapperFunction}{f_c}
\newcommand{\shortedOOPSourceInputs}{\underset{v=1\cdots k}{\langle x_v \rangle}}
\newcommand{\shortedOOPFollowUpInputs}{\underset{w=(k+1)\cdots n}{\langle x_w \rangle}}
\newcommand{\shortedOOPAllInputs}{\underset{i=1\cdots n}{\langle x_i \rangle}}
\newcommand{\shortedOOPSourceOutputs}{\underset{v=1\cdots k}{\langle \wrapperFunction(m_v,x_v) \rangle}}
\newcommand{\shortedOOPSourceOutputsEase}{\underset{v=1\cdots k}{\langle m_v(x_v) \rangle}}
\newcommand{\shortedOOPAllOutputs}{\underset{i=1\cdots n}{\langle \wrapperFunction(m_i,x_i) \rangle}}
\newcommand{\shortedOOPAllOutputsEase}{\underset{i=1\cdots n}{\langle m_i(x_i) \rangle}}

\newcommand{\detailedInvokedMethods}{\langle m_1, \cdots,  m_n \rangle} 
\newcommand{\detailedOOPSourceAndFollowupInputs}{\langle (m_1, x_1), \cdots,  (m_n, x_n) \rangle}
\newcommand{\detailedOOPSourceAndFollowupOutputs}{\langle \wrapperFunction(m_1, x_1), \cdots, \wrapperFunction(m_n, x_n)\rangle}
\newcommand{\WrapperFunctionexampleOutput}{\wrapperFunction(m, x)}

\newcommand{\PshortedOOPAllOutputs}{\underset{i=1\cdots n}{\langle P(m_i,x_i) \rangle}}
\newcommand{\PshortedOOPSourceOutputs}{\underset{s=1\cdots k}{\langle P(m_v,x_v) \rangle}}
\newcommand{\PdetailedOOPSourceAndFollowupOutputs}{\langle P(m_1, x_1), \cdots, P(m_n, x_n)\rangle}
\newcommand{\PexampleOutput}{P(m, x)}

\newcommand{\inputs}{X}
\newcommand{\outputs}{Y}
\newcommand{\metamorphicOracle}{MOracle}
\newcommand{\relation}{\mathcal{R}}
\newcommand{\outputRelation}{\relation_o}
\newcommand{\inputRelation}{\relation_i}
\newcommand{\outputRelationWithElements}{\relation_o<\sourceInput, \followUpInput, \sourceOutput, \followUpOutput>}
\newcommand{\sourceInput}{x_{s}}
\newcommand{\followUpInput}{x_{f}}
\newcommand{\sourceOutput}{y_{s}}
\newcommand{\followUpOutput}{y_{f}}

\newcommand{\sourceElementsSet}{s}
\newcommand{\inputTransformation}{\textit{transform}}
\newcommand{\inputTransformationStoF}{\followUpInput=\textit{transform}(\sourceElementsSet) (\sourceElementsSet \subseteq \sourceInput \cup \sourceOutput)}
\newcommand{\inputTransformationItoJ}{\inputJ=\textit{transform}(\sourceElementsSet)}

\newcommand{\elementSetOfOutputRelation}{E_\alpha}
\newcommand{\elementOfSource}{e_{source}}
\newcommand{\elementOfFollowUp}{e_{followUp}}
\newcommand{\exampleE}{e}
\newcommand{\exampleEone}{e_1}
\newcommand{\exampleEtwo}{e_2}
\newcommand{\assertionFunction}{f_a(\elementSetOfOutputRelation)}

\newcommand{\inputSet}{X}
\newcommand{\outputSet}{Y}

\newcommand{\inputI}{x_{1}}
\newcommand{\outputI}{y_{1}}
\newcommand{\inputJ}{x_{2}}
\newcommand{\outputJ}{y_{2}}
\newcommand{\miI}{mi_{1}}
\newcommand{\miJ}{mi_{2}}

\newcommand{\propertyOne}{\textit{P1-Method Invocations}}
\newcommand{\propertyTwo}{\textit{P2-Relation Assertion}}
\newcommand{\propertyShortOne}{\textit{P1}}
\newcommand{\propertyShortTwo}{\textit{P2}}

\newcommand{\patternOne}{\textit{A1-BoolAssert}}
\newcommand{\patternTwo}{\textit{A2-CompAssert}}

\newcommand{\inputTestCase}{\tau}
\newcommand{\methodInvocationI}{mi_{i}}
\newcommand{\methodInvocationJ}{mi_{j}}
\newcommand{\allMethodInvocations}{all\_MI}
\newcommand{\methodInvocations}{MI}
\newcommand{\CUTsetMethodInvocations}{CUT\_MI}
\newcommand{\classSigleMethodInvocations}{class\_MI}
\newcommand{\classSet}{CLASS}
\newcommand{\classSigle}{class}

\newcommand{\involvedMethodInvocations}{MI}
\newcommand{\involvedInputSet}{X'}
\newcommand{\involvedOutputSet}{Y'}

\newcommand{\relationAssertSet}{RA}
\newcommand{\relationAssertSingle}{Ra}
\newcommand{\assertSet}{A}
\newcommand{\assertSingle}{\alpha}

\newcommand{\MRInstanceSet}{MR\_INSTANCE}
\newcommand{\MRInstanceSingle}{MR\_instance}
\newcommand{\detaildedMRInstance}{\langle \assertSingle, \involvedMethodInvocations \rangle }

\newcommand{\targetMethodsInMR}{M}
\newcommand{\detaildedMRconstituents}{\langle \targetMethodsInMR, \sourceInput, \followUpInput, \inputTransformation, \sourceOutput, \followUpOutput, \assertSingle \rangle }
\newcommand{\collectClassHavingMethodInovations}{collect\_classes} 
\newcommand{\collectMethodInovations}{collect\_method\_inovations} 
\newcommand{\isInternalClass}{is\_internal\_class}
\newcommand{\collectMIofClass}{collect\_MI\_of\_class}
\newcommand{\collectAssertions}{collect\_assertions}
\newcommand{\isRelationAssertion}{is\_relation\_assertion}

\newcommand{\testCompressDecompress}{testCompressDecompress()}
\newcommand{\testCompressDecompressMR}{testCompressDecompress\_MR(byte[])}

\newcommand{\developerWrittenTestSuite}{$\mathcal{D}$}
\newcommand{\evosuiteGeneratedTestSuite}{$\mathcal{E}$}
\newcommand{\codifiedMRBasedTestSuite}{$\mathcal{C}$}
\newcommand{\cde}{$\mathcal{C}$+$\mathcal{D}$+$\mathcal{E}$}
\newcommand{\de}{$\mathcal{D}$+$\mathcal{E}$}

\newcommand{\paperSite}{https://mr-extractor.github.io/}

\newcommand{\outputOfTool}{valid codified MRs}

\newcommand{\exampleCUTName}{TextRenderer}
\newcommand{\exampleSI}{textRder}
\newcommand{\exampleFI}{boldTextRder}
\newcommand{\exampleSO}{widthNoBold}
\newcommand{\exampleFO}{widthBold}
\newcommand{\exampleName}{simulateWidth()}
\newcommand{\underlyingMR}{\mathit{\textit{IF}\ text_2=text_1.setBold()\ \textit{THEN}}\ \\ \mathit{text_1.width()\leq text_2.width()}}
\newcommand{\underlyingMRa}{\mathit{\textit{IF}\ text_2=text_1.setBold()\ \textit{THEN}}\ \mathit{text_1.width()\leq text_2.width()}}

\newcommand{\exampleMIone}{textRder.simulateWidth()}
\newcommand{\exampleMItwo}{boldTextRder.simulateWidth()}
\newcommand{\exampleinvokedMethod}{simulateWidth()}
\newcommand{\exampleIT}{boldTextRder = textRder.text.setBold()}

\newcommand{\exampleAssertion}{assertTrue(widthNoBold <= widthBold)}
\newcommand{\exampleIrrlevantAssertion}{assertTrue( boldTextRder.getText().equals("wow"))}

\newcommand{\numIdentifiedMTCs}{11,350}
\newcommand{\numPoJwithIdentifiedMTCs}{701}
\newcommand{\numRQSoundness}{RQ1}
\newcommand{\numRQQuality}{RQ2}
\newcommand{\numRQUsefulness}{RQ3}
\newcommand{\numRQUnderstandabiltiy}{RQ4}

\newcommand{\phaseOne}{MTC Discovery}
\newcommand{\phaseTwo}{MR Synthesis}
\newcommand{\phaseThree}{MR Filtering}

%% file: src/01-introduction.tex
\section{Introduction}

\congying{[2023-08-18 10:00PM] Ready to review.}
In recent years, automated test input generation has achieved significant advances ~\cite{randoop, evosuite2011,HarmanJZ15, CadarS13}. 
However, constructing test oracles is \todo{still} a major obstacle to automated test case generation. 
\textbf{Metamorphic Testing (MT)}~\cite{1998-chen-tr} has been applied to various domains as a promising approach to addressing the test \todo{oracle problem} ~\cite{2016-segura-tse}. 
MT works by employing additional test inputs when the expected output for a given input is difficult to determine. It reveals a fault if a relation (known as a \textbf{Metamorphic Relation (MR)}) between these inputs and their corresponding outputs is violated. 
\editedT{
For instance, consider a program $P(a,b,G)$ that computes the shortest path from vertex $a$ to vertex $b$ in a large undirected graph $G$. It is difficult to determine the expected output of $P(a,b,G)$, while a correct implementation of $P$ should provide the same length of paths for $P(a,b,G)$ and $P(b,a,G)$. Validating whether the outputs of $|P(a,b,G)|$ and $|P(b,a,G)|$ are equal is much easier (the associated MR is $|P(a,b,G)|=|P(b,a,G)|$).}
An advantage of MT is that an MR can serve as an oracle that is applicable to many test inputs. 
It enables automated test case generation by integrating MRs with automatically generated test inputs~\cite{2016-segura-tse}. 
However, the \todo{design} of MRs is challenging because it requires domain-specific knowledge and relies on the expertise of testers~\cite{AhlgrenBBDDGGHL21}. This hinders the wider adoption of MT~\cite{2016-segura-tse}. 


There exist studies trying to systematically explore MRs, such as identifying MRs from software specifications~\cite{chen2016metric,sun2019metric} and searching MRs using pre-defined patterns~\cite{zhou2018metamorphic,segura2018metamorphic}
However, these approaches suffer from {a low degree of automation}, i.e., they heavily rely on manual efforts to identify concrete MRs. 
On the other hand, several automatic approaches have been proposed to infer MRs for given programs, such as machine-learning-based approaches ~\cite{kanewala2013using, blasi2021memo}, search-based approaches~\cite{zhang2014search,zhang2019automatic}, and genetic-programming-based approaches~\cite{ayerdi2021generating, AyerdiTATSA22}. 
However, {wide adoption of these approaches is challenging}. 
They are designed for programs in specific domains (e.g., numerical programs~\cite{zhang2014search,zhang2019automatic}) whose input and output values exhibit certain types of relations (e.g., equivalence relations~\cite{blasi2021memo}, polynomial relations~\cite{zhang2019automatic}, or relations that follow pre-defined patterns~\cite{ayerdi2021generating}). 

\begin{figure}[!t]
\centering
\resizebox{0.8\linewidth}{!}{\includegraphics{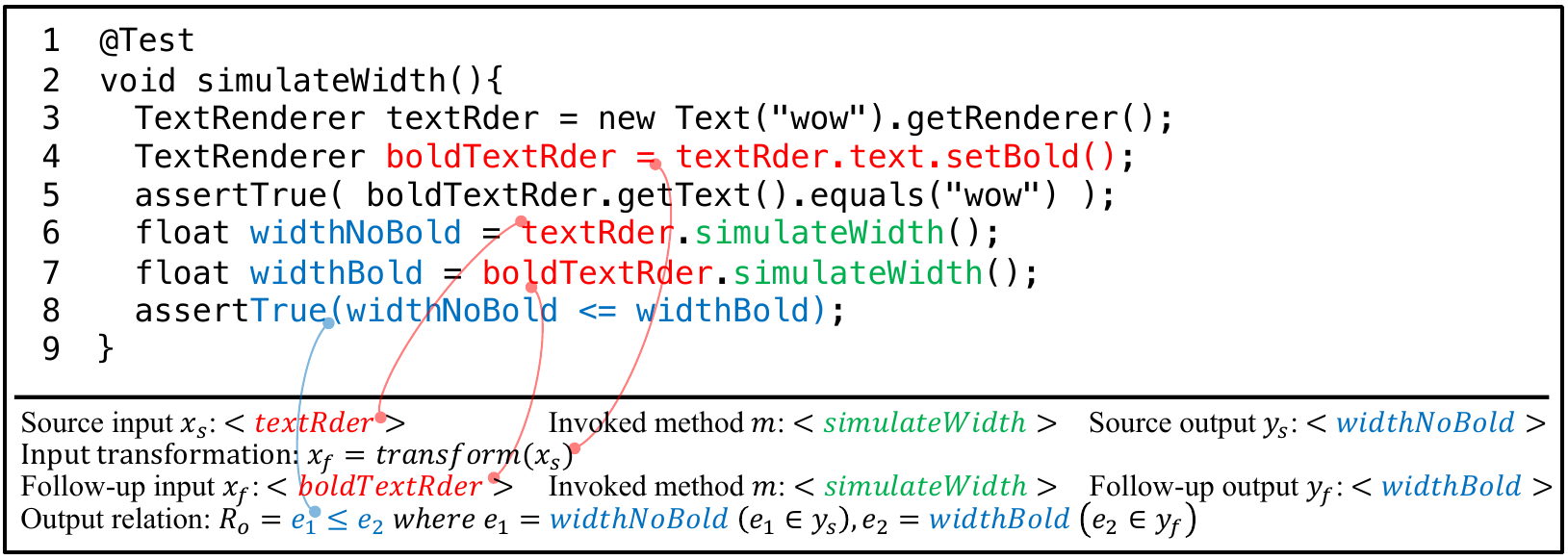}}
\caption{A test case crafted from \texttt{com.itextpdf.layout.renderer.TextRendererTest} in project \textsc{iText}. Underlying MR: $\underlyingMRa$. 
}\label{fig:embedded-MR sample-0}
\end{figure}

\textbf{Our Observations and Idea.}
We observe that the domain knowledge encoded in developer-written test cases could suggest useful MRs, even though these test cases may not originally be designed for MT. We refer to such test cases as \textit{MR-encoded test cases} (MTCs).
For example, the test case \code{\exampleName} in Figure~\ref{fig:embedded-MR sample-0} encodes the knowledge that the layout of a text should not be wider than its bold version. 
This knowledge actually suggests an MR: $\underlyingMR$. 
Moreover, these encoded MRs not only work for existing inputs (e.g., \code{Text("wow")}) but can be applicable to new inputs (e.g., \code{Text("wow!")} or \code{Text("BoldTest")}). 
This presents an opportunity of {integrating these encoded MRs with automatically generated test inputs to enable automated test case generation}~\cite{segura2018metamorphic}. 
This observation motivates us to design an automatic approach to synthesize MRs from existing test cases for automated test case generation.  


\textbf{Challenge.}
However, {automatically synthesizing MRs that are encoded in test cases presents challenges.} 
To the best of our knowledge, no existing studies have explored the discovery and synthesis of MRs from existing test cases. 
\todo{On the one hand}, there is no syntactic difference between MR-encoded test cases and non-MR-encoded test cases. 
\todo{On the other hand}, MRs are implicitly encoded in the test cases. There are no explicit indicators for the detailed constituents (e.g., source and follow-up inputs) of encoded MRs. 
For the \code{\exampleName} case in Figure~\ref{fig:embedded-MR sample-0}, there is no documentation of the encoded MR in either comments or annotations. 
After understanding the logic of this test case, we can recognize the underlying MR and its corresponding constituents.\scc{Cannot follow the argument.}
This situation presents the challenges of automatically discovering MTCs and deducing the constituents of encoded MRs. 
Consequently, to discover MRs that are encoded in test cases, our approach needs to analyze whether there is a semantic of MR in a test case. \scc{What is semantic of MR?}

\textbf{Methodology.} 
In this paper, we propose \tool, an automatic approach to \todo{discover} and synthesize MRs from existing test cases. 
To tackle the aforementioned challenges, the underlying \textbf{insight} of \tool is that {MR-encoded test cases actually comply with some properties that can be mechanically recognized}. 
Since an MR is defined over at least two inputs and corresponding outputs, we derive two principal properties that characterize an MR-encoded test case --- \todo{(i) containing executions of target programs on at least two inputs, and (ii) containing the validation of a relation over these inputs and corresponding outputs.} 

Specifically, \tool works in three phases. 
\tool first discovers MTCs based on the two derived properties (Section~\ref{sec:app-phase1}, \emph{MTC Discovery}). 
Then, with discovered MTCs, \tool deduces the constituents (e.g., source and follow-up inputs) of encoded MRs and then codifies these constituents into parameterized methods to facilitate automated test case generation. These parameterized methods are termed \editedTS{\textit{codified MRs}}  (Section~\ref{sec:app-phase2}, \emph{MR Synthesis}). 
\editedTS{Finally, \tool filters out codified MRs that demonstrate poor quality in applying to new test inputs. This is because codified MRs that are not applicable to new test inputs are {useless} for new test generation (Section~\ref{sec:app-phase3}, \emph{MR Filtering}).}


\textbf{Evaluation.} 
We built a dataset of over 11,000 MTCs discovered by \tool from 701 OSS projects in the wild. 
To evaluate the precision of \tool in discovering MTCs, \editedTS{we manually examined 164 randomly selected samples}, and found 97\% of them are true positives \editedT{that satisfy the defined properties of an MTC.} This indicates the high precision of \tool in discovering MTCs and the high reliability of our MTC dataset (Section~\ref{sec:rq-soundness}, \numRQSoundness). 
\tool synthesizes codified MRs from MTCs and applies filtering to remove low-quality MRs. To evaluate the effectiveness of this process, 
\editedT{we employed EvoSuite to automatically generate a set of new test inputs for each codified MR}. Experimental results show that \todo{97.18\%} of codified MRs are of high quality and applicability to new inputs for automated test case generation, demonstrating the practical applicability of \tool (Section~\ref{sec:rq-quality}, \numRQQuality). 
Furthermore, to demonstrate the usefulness of \todo{synthesized MRs in complementing existing tests and enhancing test adequacy}, we compared test suites constructed from codified MRs against developer-written and EvoSuite-generated test suites. 
Experimental results show \todo{13.52\%} and \todo{9.42\%} increases in the line coverage and mutation score, respectively, when the developer-written test suites are augmented with codified-MR-based test suites. 
As to EvoSuite-generated test suites, there is an \todo{82.8\%} increase in mutation score (Section~\ref{sec:rq-usefulness}, \numRQUsefulness). 
To evaluate the comprehensibility of codified MRs, we conducted a qualitative study involving \todo{five} participants and \todo{52} samples. Results show that \todo{55.76\% to 76.92\%} of codified MRs are easily comprehended, \todo{showcasing their potential for practical adoption by developers}. 

\textbf{Contribution.} Our work makes the following contributions.
\begin{itemize}
    \item We propose \tool, the first approach that automatically synthesizes MRs from existing test cases.
    \item We release a dataset of over 11,000 MTCs discovered across \numPoJwithIdentifiedMTCs{} OSS projects, and investigate their distribution and complexity. This dataset stands as a valuable resource for future research in fields such as MR discovery, MR inference, and automated MT. 
    \item We conduct extensive experiments to evaluate the precision of \tool in discovering MTCs and evaluate the quality, usefulness, and \todo{comprehensibility} of MRs synthesized by \tool.
    \item We release the research artifact and all experimental datasets on \tool's website~\cite{mr-extractor} to facilitate reproducing our experimental results and future research.
\end{itemize}

%% file: src/02-preliminary.tex

\section{Preliminaries}
\congying{[August 11, 2023] This section is revised and ready for you to review.}\scc{The writing is a bit rough. It takes me quite some time to go over one subsection. Please consider improving the writing. I expect that the writing can be in better shape before my review.}
\congying{Professor, thanks for your time. I am trying my best to write as well as possible. I have learned a lot from Hengcheng's and Jiarong's comments and revisions. Hope I can do better.
}

\subsection{Metamorphic Testing}
Metamorphic testing is a process that tests a program $\programUnderTest$ with a metamorphic relation.
Given a sequence of inputs (\textbf{source inputs}) and their program outputs (\textbf{source outputs}), additional inputs (\textbf{follow-up inputs}) are constructed to obtain additional program outputs (\textbf{follow-up outputs}). 
If these inputs and outputs do not satisfy the metamorphic relation, $\programUnderTest$ contains a fault. 

\textbf{Metamorphic Relation (MR).}
Let \emph{f} be a target function. 
A metamorphic relation of \emph{f} is a property defined over a sequence of inputs $\detailedSourceAndFollowupInputs$ ($n \geq 2$) and their corresponding outputs $\detailedSourceAndFollowupOutputs$~\cite{chen2018metamorphic}. 
Following the definition by Segura et al.~\cite{SeguraDTC17}, an MR can be formulated as a logical implication from an \textbf{input relation} $\inputRelation$ to an \textbf{output relation} $\outputRelation$. 
\[ \relation_i \left( \shortedSourceInputs, \shortedFollowUpInputs, \shortedSourceOutputs \right) \implies \outputRelation \left( \shortedAllInputs, \shortedAllOutputs \right) \] 

$\relation_i$ is a relation over source inputs $\detailedSourceInputs$, follow-up inputs $\detailedFollowUpInputs$, and source outputs $\detailedSourceOutputs$. 
\todo{The inclusion of source outputs in $\inputRelation$ allows follow-up inputs to be constructed based on both source inputs and outputs.}
$\relation_o$ is a relation over all inputs $\detailedSourceAndFollowupInputs$ and the corresponding outputs $\detailedSourceAndFollowupOutputs$. 
The MR formulation is a general form of that proposed by Chen et al.~\cite{1998-chen-tr}. 
It expresses an MR in terms of an input relation and an output relation.
\begin{exmp}
\editedT{
Consider a function $f(a,b,G)$ computing the shortest path from vertex $a$ to vertex $b$ in an undirected graph $G$.
    The property $|f(a,b,G)|=|f(b,a,G)|$ implies that the length of the shortest path should be the same in either direction ($a$ to $b$ or $b$ to $a$), and it can be formulated as
    \[
        x_2 = t(x_1) \implies |f(x_2)| = |f(x_1)| \ \text{where}\ t((a, b, G)) = (b, a, G)
    \]
    In this case, \(\mathcal{R}_i = \left\{ \left((v_1, v_2, G), (v_2, v_1, G)\right) \mid \forall G, \forall v_1, v_2 \in G\right\}\) includes all pairs of inputs to \(f\) such that the first two elements (source and sink vertexes) are swapped. \(\mathcal{R}_o = \left\{ (n, n) \mid \forall n \in \mathbb{N} \right\}\) includes all pairs of equivalent numbers (shortest paths).
}
\end{exmp}

\textbf{Metamorphic Testing (MT).}\label{sec:Metamorphic Testing}
Given an MR $\relation$ for a function $f$,  metamorphic testing is the process of validating $\relation$ on an implementation $\programUnderTest$ of $f$ using various inputs~\cite{SeguraDTC17}.

Intuitively, assuming a program implemented by a sequence of statements, MT entails the following \todo{five} steps~\cite{chen2018metamorphic}:
(i) constructing a source input, which can be written by developers or automatically generated (e.g., random testing)~\cite{2016-segura-tse},
(ii) executing the program with the source input to get the source output,
(iii) constructing a follow-up input that satisfies $R_i$,
(iv) executing the program with the follow-up input to get the follow-up output,
and (v) checking if these inputs and outputs satisfy the output relation $\outputRelation$.

In MT, the input relation $\relation_i$ is used for constructing the test inputs in the first three steps. 
Typically, a function, referred to as \textbf{input transformation}, is designed to construct a follow-up input satisfying $\inputRelation$ from a source input and/or source output. The output relation $\outputRelation$ serves as the oracle in the last step. 
For example, in Figure~\ref{fig:embedded-MR sample-0}, the statement \code{boldTextRder = textRder.text.setBold()} transforms the source input \code{textRder} to the follow-up input \code{boldTextRder}, and 
the output relation \code{assertTrue(widthNoBold <= widthBold)} gives the oracle.

\subsection{\editedA{Adaptation of MR Formulation in the Context of OOP}}\label{sec:MROOPFormulation}
\begin{figure}[t]
\centering
\resizebox{0.72\linewidth}{!}{\includegraphics{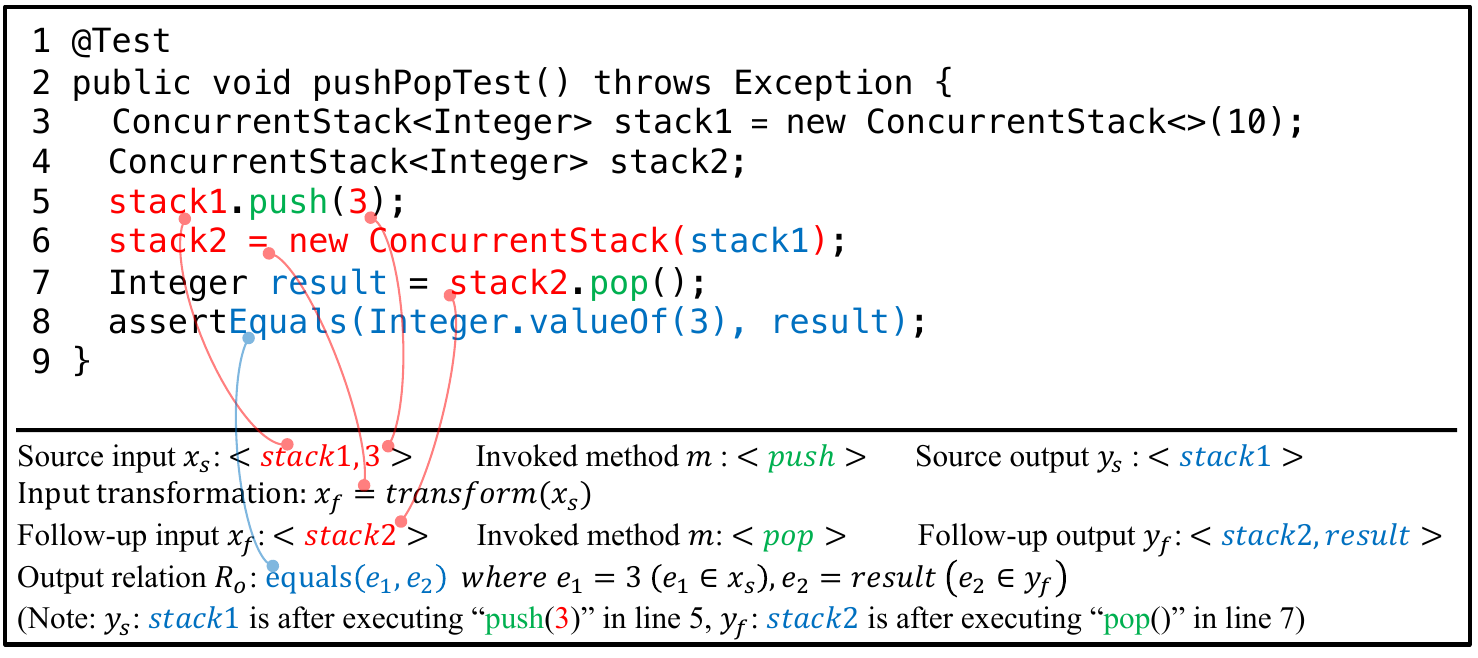}}
\caption{A test case crafted from \code{com.conversantmedia.util.concurrent.ConcurrentStackTest} in project \textsc{Disruptor}. Underlying MR: $\mathit{x=stack.push(x).pop()}$ --- \textit{IF} an element $x$ is pushed onto a stack and the stack subsequently pops off the top element, \textit{THEN} the element $x$ should be the one popped.
}\label{fig:embedded-MR sample-1} 
\end{figure}

Given the observation that developers encode MRs in test cases as oracles (as exemplified in Figure~\ref{fig:embedded-MR sample-0}), our goal is to automatically discover and synthesize these encoded MRs from existing test cases in open-source projects. 
This paper focuses on unit test cases for object-oriented programming (OOP) programs. 
Since the existing MR formulation is not originally designed for OOP programs, we make a slight adaptation. Specifically, a unit under test refers to a ``class'' rather than a single function ($\targetFunction$) in MR formalism. 
Therefore, a unit test case for a class under test (CUT) can comprise more than one method invocation. 
\todo{It implies that a metamorphic relation for a class may involve more than one function.} 
For example, in Figure~\ref{fig:embedded-MR sample-1}, the underlying relation $x=stack.push(x).pop()$ is over two functions \code{push} and \code{pop} from a stack class. 

\begin{figure}[!t]
\centering
\resizebox{0.8\linewidth}{!}{\includegraphics{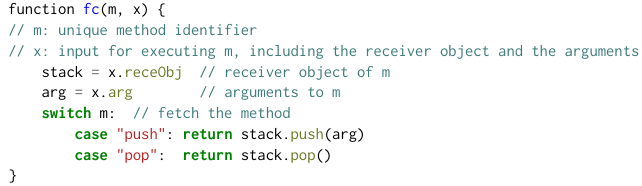}}
\caption{Illustration of a wrapper function $\wrapperFunction$ for a stack class implemented with methods \code{push} and \code{pop}.
(The output of \code{fc("push",x)} is a \code{stack} object which has just pushed \code{arg} into it, while the output of \code{fc("pop",x)} are the popped element by executing \code{stack.pop()} \todo{and the \code{stack} object which has just popped an element}.)
}
\label{lst:wrapper-function}
\end{figure}

To accommodate this, we ``wrap'' the semantics of a class (including its methods) by a function called \textbf{class wrapper function} $\wrapperFunction$. 
$\wrapperFunction$ takes as input a method identifier $m$ and the input $x$ for $m$, and then invokes $m(x)$ internally. 
Figure~\ref{lst:wrapper-function} presents an illustration of $\wrapperFunction$ wrapping a stack class with methods \code{push} and \code{pop}.
As a result, we can formulate an MR for the stack class based on a single wrapper function instead of functions \code{push} and \code{pop}.
\valerio{it is not very clear here, first of all we are not mentioning the MR related to push and pop, second, it is not very clear to a reader. I understand the MT is explained later but should be explained before, just discuss the test case in FIgure 2 first, and then show the example of wrapper function with the MR }
\congying{already added.}


Let $\WrapperFunctionexampleOutput$ denote the output of $\wrapperFunction$ invoking the method $m$ on the input $x$.
An MR $\relation$ over a sequence of inputs $\detailedSourceAndFollowupInputs$ $(n \geq 2)$ with additional corresponding method identifiers $\detailedInvokedMethods$ and their corresponding outputs $\detailedOOPSourceAndFollowupOutputs$ can be formulated as follows.

\begin{align}
\relation_i \Big( \shortedOOPSourceInputs, \shortedOOPFollowUpInputs,  \shortedOOPSourceOutputs \Big) \nonumber  \implies \outputRelation \Big( \shortedOOPAllInputs, \shortedOOPAllOutputs \Big) \nonumber
\end{align}


For ease of presentation, in the remainder of the paper, we use $m(x)$ to denote $f_c(m, x)$, where $m$ is the delegated method in the class under test. 

\begin{exmp}
Let $\wrapperFunction$ stand for the class under test \code{ConcurrentStack} in Figure~\ref{fig:embedded-MR sample-1}. 
Given the illustration in Figure~\ref{lst:wrapper-function}, the relation 
$\mathit{x=stack.push(x).pop()}$ can be expressed as: \textit{IF} two inputs $\langle x_1, x_2 \rangle$ have the relation $\mathit{x_2.receObj= push(x_1)}$ $(\relation_i)$ , \textit{THEN} the output relation $\mathit{pop(x_2) = x_1.arg}$ $(\outputRelation)$ is expected to be satisfied. 

In this test case, $\mathit{x_1.receObj}$ and $\mathit{x_1.arg}$ are implemented with \code{stack1} and \code{3}, and the invocation $\mathit{push(x_1)}$ is implemented as \code{stack1.push(3)}. 
Similarly, $\mathit{x_2.receObj}$ and $\mathit{pop(x_2)}$ are implemented with \code{stack2} and \code{stack2.pop()} (\code{pop()} does not require any argument). 
The expected relation $\mathit{pop(x_2) = x_1.arg}$ is validated by \code{assertEquals(Integer.valueOf(3), result)}.
\end{exmp}

\begin{exmp}
When the function $\wrapperFunction$ wraps the class \code{\exampleCUTName} in Figure~\ref{fig:embedded-MR sample-0}, the relation $\underlyingMRa$ can be expressed as: \textit{IF} two inputs $\langle x_1, x_2 \rangle$ have the relation $\mathit{x_2.receObj\ = x_1.receObj.text.setBold()}$ ($\relation_i$) , 
\textit{THEN} the output relation $\mathit{simulateWidth(x_1) \leq simulateWidth(x_2)}$ ($\outputRelation$) is expected to be satisfied. 

In this test case, $\mathit{x_1.receObj}$ and $\mathit{x_2.receObj}$ are implemented with \code{textRder} and \code{boldTextRder}. Arguments are not needed for \code{simulateWidth()}, i.e., $\mathit{x_1.arg=x_2.arg=null}$. 
The statement \code{assert\\True(widthNoBold <= widthBold)} validates whether the execution results of \code{textRder.simulate\\Width()} and \code{boldTextRder.simulateWidth()} satisfy the expected output relation $\outputRelation$. 
\end{exmp}

\editedT{
Note that, for methods with nested method calls, such as $m(g(\cdot))$,
the nested method call $g(\cdot)$ is considered as an argument to the method $m$. $m(g(\cdot))$ can be expressed as $m(x)$ where $\mathit{x.arg=g(\cdot)}$.
For methods having side effects, the identification of such methods' outputs will be discussed in Section~\ref{sec:app-phase1-elementofrelation}. 
} 


\begin{figure*}[!t]
\centering
\resizebox{\linewidth}{!}{\includegraphics{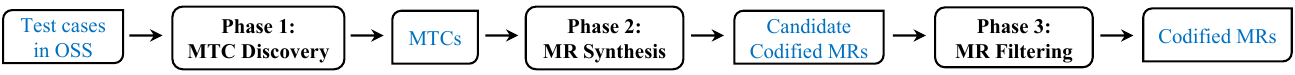}}
\caption{Overview of \tool}\label{fig:overview}
\end{figure*}

%% file: src/03-approach.tex
\section{Methodology}
\congying{[August 18, 2023] This section is ready for you to revise.}

Inspired by our observation that test cases written by developers can embed domain knowledge that encodes MRs, we propose an approach, \tool{}, to discover and synthesize encoded MRs from existing test cases automatically. 
The underlying insight of \tool is that encoded MRs obey certain semantic properties that can be mechanically recognized. 
Figure \ref{fig:overview} presents an overview of \tool.
\tool takes as input test cases collected from open-source projects and returns codified MRs.
Specifically, \tool works in the following three phases.
\begin{enumerate}
\item \textit{\phaseOne}.
    According to the formulation of MR, we derive two principal properties that characterize an MR-encoded test case. 
    First, the test case must contain at least two invocations to methods of the same class with two inputs separately (P1). 
    Second, the test case must contain at least one assertion that validates the relation between the inputs and outputs of the above method invocations (P2). 
    This is because an MR is defined over at least two inputs and corresponding outputs.
    These two properties guarantee the execution of at least two inputs and the validation of the output relation over these inputs and outputs. 
    By checking the above properties, \tool can mechanically discover MR-encoded test cases (MTCs) in open-source projects (Section~\ref{sec:app-phase1}).
\item \textit{\phaseTwo.}
    Given MR-encoded test cases and corresponding method invocations and relation assertions identified in the \textit{\phaseOne} phase, \tool first deduces the MR constituents (e.g., source input and follow-up input) and then codifies their constituents into parameterized methods \todo{to facilitate automated test case generation}. 
    Such methods are termed \textit{codified MRs} in this paper (Section \ref{sec:app-phase2}). 
\item \textit{\phaseThree.}
    We target discovering MRs for new test generation. Codified MRs not applicable to new test inputs are ineffective for new test generation~\cite{qiu2020theoretical}. Therefore, in this phase, \tool filters out codified MRs that demonstrate poor quality \todo{(e.g., leading to false alarms)} in applying to new source inputs.
\end{enumerate}

\begin{figure*}[!t]
\vspace{3pt}
\centering
\resizebox{\linewidth}{!}{\includegraphics{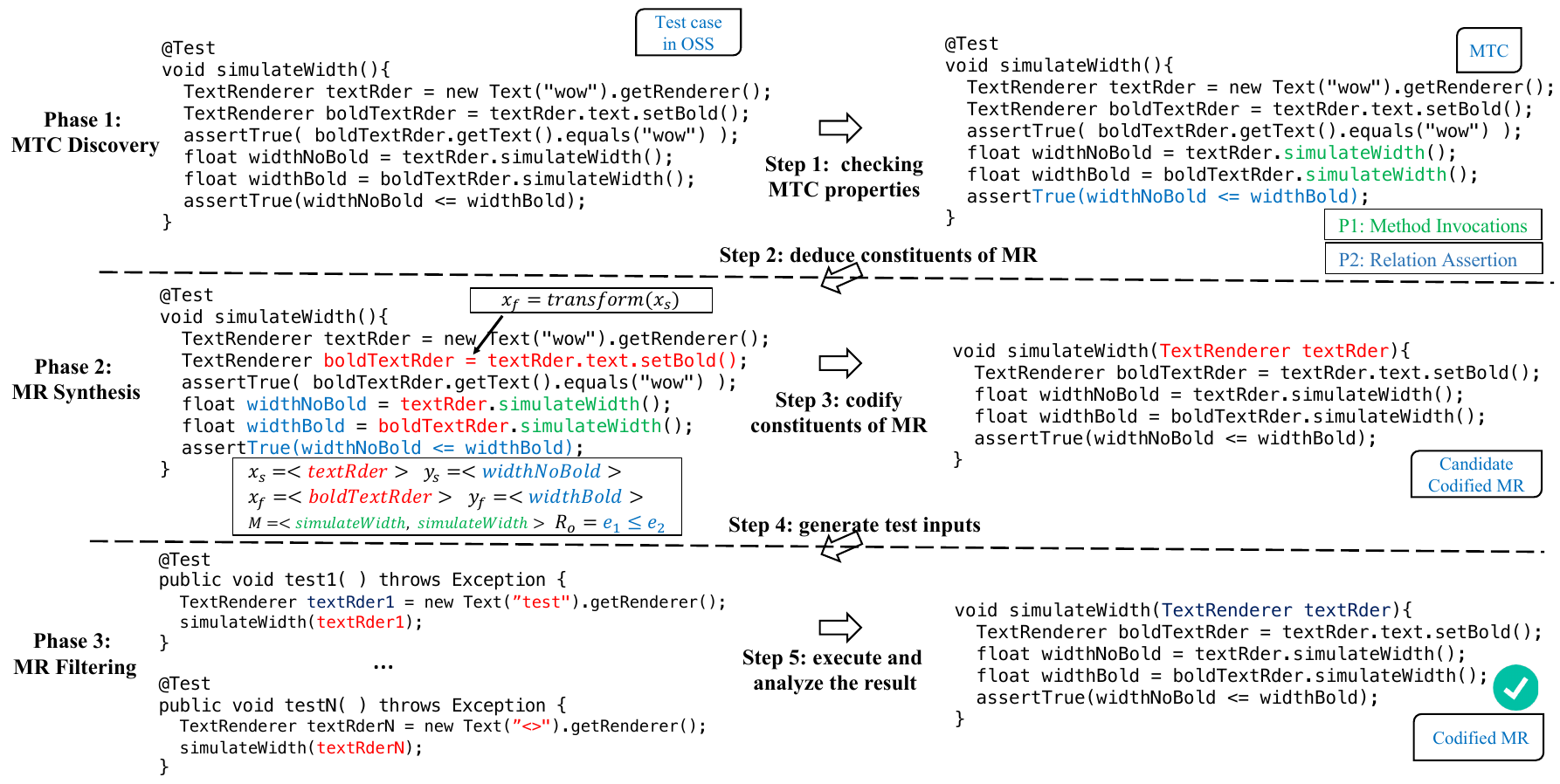}}
\caption{Procedure of \tool operating on the MTC \code{simulateWidth()}
}\label{fig:procedure}
\end{figure*}

\begin{table*}[t]
\centering
\footnotesize
\caption{Assertion APIs and examples for relation assertions patterns}\label{table:assertion method name}
\input{table/relation-assertion-pattern}
\end{table*}

\subsection{Phase~1: MTC Discovery}\label{sec:app-phase1}
Phase~1 of \tool aims to discover MR-encoded test cases (MTCs). 
Unfortunately, MTCs are not explicitly labeled and have no syntactic difference with test cases without MRs.
Therefore, to discover possible MTCs, \tool should analyze whether the given test cases embed the semantics of an MR. 
So we first model the semantics of an MR-encoded test case with two principal properties that can be mechanically analyzed. 
Then, \tool checks these properties in given test cases from open-source projects. 
\todo{Test cases that satisfy the two properties are considered as MTCs by \tool.} \label{sec:criteria-MTC}

\subsubsection{\editedT{Properties of An MTC}} \label{sec:necessary properties of MT}
According to the formulation of MR in Section \ref{sec:Metamorphic Testing}, we derive two properties (\propertyOne{} and \propertyTwo{}) of an MTC.

\begin{enumerate}
\item[\textbf{P1}] \textbf{Method Invocations}: 
\emph{The test case should contain \underline{at least two invocations} to the methods of the same class \todo{with two inputs separately}. }
This class is considered as a class under test, and the method invocations are denoted as $\mathit{\methodInvocations}$. 
This property is derived from the fact that an MR is defined over at least two inputs and corresponding outputs.
\editedT{When there are at least two method invocations and each method invocation has a pair of input and corresponding output, this ensures the existence of at least two inputs and corresponding outputs.} 
Specifically, we allow the invocations to the same or different methods of a class under test.


\item[\textbf{P2}] \textbf{Relation Assertion}: 
\emph{The test case should contain \underline{at least one assertion} checking the relation between the inputs and outputs of the invocations in \(MI\).} 
This property is derived from the fact that an MR has a constraint (i.e., $\outputRelation$) over the input and outputs of program executions (i.e., method invocations). Such an assertion is to validate the output relation $\outputRelation$. 
\end{enumerate}

\subsubsection{Step 1: Checking MTC properties}
When checking \propertyOne{}, \editedT{\tool first collects all the method call sites within a test case, and focuses on methods from internal classes that are native to the project under analysis}.
\tool excludes methods of external classes (such as a class from a third-party library) that are not target classes to test, by matching the prefix of their fully qualified names~\cite{DBLP:conf/kbse/ZhuWWLCSZ20}.
For each internal class with at least two method invocations, the class is considered as a class under test. 
All classes under test and corresponding method invocations are collected to facilitate \propertyTwo{} identification. 

However, when it comes to checking \propertyTwo{}, \tool encounters a technical issue: \textit{how to automatically distinguish output relations that are implicitly encoded in assertion statements.} 
It can be difficult to tell whether an assertion statement represents a genuine relation over multiple outputs or simply a combination of separate output assertions for convenience. 
For instance, consider an assertion statement with two outputs \code{y1} and \code{y2} (e.g., \code{assertTrue(y1==-1 \&\& y2==1)}).
It is ambiguous whether the relation ``y1==-y2'' should hold or it is a shortcut for \code{assertTrue(y1==-1)} and \code{assertTrue(y2==1)}.

To deal with the above issue, we propose two general assertion patterns where an output relation can be modeled and validated.
Assertions matching these patterns are considered checking an output relation.
The design principle of the two patterns is that \textit{an output relation is essentially a boolean expression that relates elements (i.e., inputs and outputs) of method invocations}. We first introduce the necessary elements of an output relation, and then introduce how these elements should be related.

\textbf{\editedT{Necessary} Elements of An Output Relation.} \label{sec:app-phase1-elementofrelation}
According to the formulation of MR, an output relation is defined over a set of inputs and outputs. 
However, there are constraints: (i) the inclusion of a follow-up output, and (ii) the inclusion of either a source output or a source input. 
As to constraint-(i), the absence of a follow-up output suggests that the second method invocation is not required. This contradicts the definition of MT which requires at least two method invocations. As to constraint-(ii), the absence of a source output and a source input suggests that the first method invocation is not required, contradicting the definition of MT. 
Note that an output relation can be defined only over a follow-up output and a source output, as illustrated in Figure~\ref{fig:embedded-MR sample-0} where the follow-up output (\code{widthBold}) and source output (\code{widthNoBold}) are included in the output relation.
Alternatively, an output relation can be defined only over a follow-up output and a source input. For the case in Figure~\ref{fig:embedded-MR sample-1}, the follow-up output (\code{result}) and source input (\code{3}) are included in the output relation.

Given method invocations $\mathit{MI}$=$\mathit{\{mi_i\}_{i=1}^n}$ of a class under test, if an assertion $\assertSingle$ is verifying an output relation, $\assertSingle$ must have two elements (donated as $e_1$ and $e_2$). 
$e_1$ is the input or the output of a method invocation ($\mathit{mi_1}$), and $e_2$ is the output of another method invocation ($\mathit{mi_2}$) that is invoked after $\mathit{mi_1}$. This allows $e_1$ to be the source input or output and $e_2$ to be the follow-up output, satisfying the above two constraints respectively. 

Next, we discuss what are the input $x_i$ and output $y_i$ of a method invocation $\mathit{mi_i}$. 
According to the specification of Java~\cite{JLS}, method invocation $\mathit{mi_i}$ can be presented as $\mathit{returnV=receObj.m_i(arg)}$, where $\mathit{returnV}$ represents the return value of the invoked method $\mathit{m_i}$, $\mathit{receObj}$ represents the receiver object of method $m_i$, and $\mathit{arg}$ represents the input parameter for executing $m_i$. 
\begin{itemize}
    \item Input $x_i$ is a set, including (i) the input arguments $\mathit{arg}$ (primitive values or object references) and (ii) the receiver object $\mathit{receObj}$ (if its fields are accessed in the method invocation). 
    \hc{How do we tell if the fields are accessed? Did we conduct static analysis on the CUT?}
    \hc{Can we always include the receiver?}
    \item Output $y_i$ is a set, including (i) the return value $\mathit{returnV}$ (if any), (ii) the receiver object $\mathit{receObj}$ after the method invocation (if the receiver object's field is updated during the method invocation), and (iii) the objects in $\mathit{arg}$ after the method invocation (if these input objects' fields are updated during the method invocation). 
    \hc{Can we always include the receiver and the arguments? It is fine that they are neither accessed nor updated, but it can simplify our definition.}
\end{itemize}

\begin{exmp}
For the test case in Figure~\ref{fig:embedded-MR sample-1}, there are two method invocation $\mathit{mi_1=}$ \code{stack1.push(3)} on line \todo{5} and $\mathit{mi_2=} $\code{stack2.pop()} on line \todo{7}, where $x_1=\{ \code{stack1}, \code{3}\}$, $x_2=\{\code{stack2}\}$, $y_1=\{\code{stack1}\}$ \text{(just after \code{stack1.push(3)})}, and $y_2=\{\code{result}, \code{stack2}\}$ \text{(just after \code{stack2.pop()})}. 
The assertion $\assertSingle$ on \todo{line 8} can be interpreted as $\code{assertEquals}$ $\code{(Integer.valueOf(} e_1\code{),}e_2\code{)}$, where $e_1=$ $\code{3}$ ($e_1 \in x_1$) and $e_2=$ $\code{result}$ ($e_2 \in y_2$), satisfying the above constraint of elements in an output relation.
\end{exmp}

\textbf{Patterns of Relation Assertions.} 
In addition to the above constraint telling if an assertion includes necessary elements ($\exampleEone$ and $\exampleEtwo$) of an output relation, we further check if $\exampleEone$ and $\exampleEtwo$ are related by a boolean expression\hc{the second pattern is not a boolean expression} with the following two patterns.

\todo{Inspired by existing work on synthesizing assertion oracles with a set of boolean and numerical operators~\cite{valerioJTP20},} the principle of two patterns is that an output relation assertion should be a boolean expression where necessary elements $\exampleEone$ and $\exampleEtwo$ are related by (i) numerical operators or user-defined boolean methods (\patternOne) or (ii) assertion methods provided by testing frameworks (\patternTwo). 
\hc{Can we call them unitary assert and binary assert?}

\begin{enumerate}
    \item[\textbf{A1}] \textbf{BoolAssert}: For assertions with a boolean parameter, such as \code{assertTrue}, $\exampleEone$ and $\exampleEtwo$ should be related by (i) numerical operators (i.e., $=, <, >, \leq , \geq, \neq$), or (ii) user-defined methods that return boolean values.
    \begin{exmp}
    The assertion \code{\exampleAssertion} in Figure~\ref{fig:embedded-MR sample-0} can be mapped onto \patternOne{} , where $\exampleEone=$ \code{\exampleSO} and $\exampleEtwo=$ \code{\exampleFO} are related by a numerical operator ``$\leq$''.
    \end{exmp}
    
    \item[\textbf{A2}] \textbf{CompAssert}:  For assertions with parameters for comparison, such as \code{assertEquals}, one of the parameters should contain $\exampleEone$, and the other should contain $\exampleEtwo$. 
    \begin{exmp}
    The assertion \code{assertEquals(Integer.valueOf(3), result)} in Figure~\ref{fig:embedded-MR sample-1} can be mapped onto \patternTwo, where $\exampleEone=$ \code{3} and $\exampleEtwo=$ \code{result}. $\exampleEone$ and $\exampleEtwo$ are related by the method \code{Arrays.equals} which returns a boolean result.
    \end{exmp}
\end{enumerate} 

The above two patterns can cover the most commonly used assertion APIs. In Table~\ref{table:assertion method name}, the corresponding APIs in \textsc{JUnit4}~\cite{junit4-assert} and \textsc{JUnit5}~\cite{junit5-assert} and some abstract examples of the above two patterns are presented. 
Assertions that match the two patterns are considered to validate an output relation. 
It should be noted that there is a trade-off between precision and completeness in recognizing relation assertions. 
In order to recognize relation assertion precisely, \todo{our patterns exclude elements related by logical operators, such as AND, OR, XOR, and EXOR. This is because elements related by these logical operators may not inherently denote a relationship. 
For example, an assertion \code{assertTrue(y1 \&\& y2)} can be merely a combination of \code{assertTrue(y1)} and \code{assertTrue(y2)} for convenience, without an actual output relation between \code{y1} and \code{y2}. 
While excluding logical operators may cause \tool to miss some output relations, reducing the risk of confusing or misleading developers with incorrect MRs is pretty important. 
}
\hc{We need to discuss what kind of of CORRECT relation assertions would be missed with an example as well as how we mitigate this threat.}

In a test case, assertions fitting into the above two patterns are considered relation assertions. 
This indicates that this test case satisfies \propertyTwo{} and is discovered as an MTC. 
Note that developers may encode more than one MR in a single test case. We consider the application of an MR for a specific set of inputs and outputs as an \textbf{MR instance}. In MTCs, an MR instance is implemented by a relation assertion over the inputs and outputs of method invocations of a class under test. 
An MR instance in an MTC is denoted as a tuple $\mathit{\langle \assertSingle, \methodInvocations \rangle}$, where $\assertSingle$ denotes a relation assertion and $\mathit{\methodInvocations}$ denotes corresponding method invocations whose output relation is validated by $\assertSingle$. 
\tool collects all MR instances in an MTC to facilitate the following MR synthesis. 



\subsubsection{\editedT{Detailed Analysis Process and Limitations}}
\tool~\cite{mr-extractor} is \todo{implemented} to statically analyze the source code of test cases. 
In checking \propertyOne{}, \tool initially collects all method invocations within a given test case.
By analyzing the fully qualified names of method invocations, \tool identifies and collects internal classes with at least two method invocations as classes under test. 
The presence of at least two method invocations in a single internal class indicates that \propertyShortOne{} is satisfied. 
However, the static analysis of source code may cause imprecise results. For example, the fully qualified names of invoked methods might be wrongly identified if overriding exists. 
This leads to the satisfaction of \propertyOne{} being wrongly detected. 

As to \propertyTwo{}, \tool collects all assertion statements in the given test case. Then, for each assertion, \tool checks whether this assertion is checking the output relation over the inputs and/or outputs of method invocations which are identified in \propertyShortOne{}. 
During the identification of inputs and output of a method invocation, to tell whether a receiver object is an input and/or an output, \tool analyzes the method's call chain and analyzes whether the fields of the object are accessed or updated in each method of the call chain. 
However, factors such as aliasing and path sensitivity present challenges in precisely analyzing whether the object's fields are accessed or updated. 

\subsection{Phase~2: MR Synthesis}\label{sec:app-phase2} 


\congying{[2023-08-18 2:30 PM] This paragraph about ``non-sufficient condition for an MTC'' is very important and dangerous. Please help me check whether it is well presented.} 
With discovered MTCs in Phase~1, in this phase, \tool synthesizes MRs from these MTCs.
\todo{
However, this process is not straightforward since some encoded MRs are incomplete. 
Properties \propertyOne{} and \propertyTwo{}, while being principal and necessary, only concern the output relation ($\outputRelation$) of an MR.
Albeit an MR is applied and validated in a test case, 
the input relation ($\inputRelation$) can be \todo{implicit} or even absent. Specifically, for MTCs where the inputs are hard-coded, the input relation is unclear. Inferring the potential relation between hard-coded values is a challenging problem. To the best of our knowledge, no existing study explores this problem, nor does our paper. In this paper, we focus on synthesizing MRs from MTCs where input relations are explicitly encoded, i.e., having input transformation that constructs follow-up inputs satisfying $\inputRelation$ from source inputs and/or source outputs.
}


The synthesis process involves (i) deducing the constituents of an encoded MR and (ii) codifying these constituents into an executable method that is parameterized with source inputs. This parameterized method is referred to as \textit{codified MR}. 
By making these methods parameterized with source inputs, new values of source inputs can be easily generated by \todo{automatic tools} (e.g., Randoop~\cite{randoop} and EvoSuite~\cite{evosuite2011}) and utilized for automated test case generation. 
These codified MRs are composed of (i) an input transformation, (ii) executions of source and follow-up inputs, and (iii) an output relation assertion.


\subsubsection{\editedT{Step 2: Deducing Constituents of an MR Instance}}
Developers may encode multiple MRs in a single test case, where a set of MR instances can be identified from an MTC. 
Following the notations in Phase~1 (Section~\ref{sec:app-phase1-elementofrelation}), for each MR instance $\mathit{\detaildedMRInstance}$, 
\tool deduces a tuple of detailed constituents, including (1) the \textbf{target methods}, (2) the \textbf{source input}, \textbf{follow-up input}, and the \textbf{input transformation}, and (3) the \textbf{source output}, \textbf{follow-up output} and the \textbf{output relation assertion}. The details of the deduction are as follows. 

\begin{enumerate}
\item[(1)] \tool takes methods invoked in $\mathit{\methodInvocations}$ as target methods. 
\item[(2)] As to input-related constituents, \tool first identifies the existence of transformation $\mathit{\inputTransformationItoJ}$ $\mathit{(\sourceElementsSet \subseteq \inputI \cup \outputI)}$, where $\sourceElementsSet$ is the input $\inputI$ and/or output $\outputI$ of a method invocation $\miI$, and $\inputJ$ is the input of $\miJ$ ($\mathit{\miI,\miJ \in \methodInvocations}$). 
Then, \tool takes $\inputI$ and $\inputJ$ as the source input and the follow-up input, respectively. 

Note that not all MR instances have the input transformation because follow-up inputs can be hard-coded rather than constructed from the source inputs and the source outputs. 
This paper only focuses on MRs with input transformation. 
Besides, \tool synthesizes MRs from MR instances that involve exactly two method invocations ($\mathit{|\involvedMethodInvocations|=2}$). This is similar to existing studies~\cite{zhou2018metamorphic,2016-segura-tse,chen2016metric,chen2018metamorphic}. 
Our evaluation results reveal that \todo{64.13\%} of MR instances only involve two method invocations (Section \ref{sec:rq-prevalence}), indicating that \tool can deal with a large portion of MR instances. Synthesizing MRs from instances that involve more than two method invocations can be challenging and interesting future work.

\item[(3)] As to output-related constituents,
\tool directly takes the source input corresponding output as the source output, takes the follow-up input corresponding output as the follow-up output, and takes the output relation assertion $\assertSingle{}$ in the identified MR instance. 
\end{enumerate} 

\begin{exmp}
In Phase 2 of Figure \ref{fig:procedure}, there is only one MR instance where the output relation assertion $\assertSingle$ is \code{\small\exampleAssertion}
and the method invocations $\mathit{MI}$ are $\mathit{\langle\small\code{textRder.simulateWidth()},\small\code{boldTextRder.simulat}\code{eWidth()}\rangle}$. \\
The identified target method is \code{\small\exampleinvokedMethod}, the source input is \code{\small\exampleSI}, the follow-up input is \code{\small\exampleFI}, the input transformation is {\code{\small boldTextRder = textRder.text.setBold()}}, the source output is \code{\small\exampleSO}, the follow-up output is \code{\small\exampleFO}, and the output relation assertion $\assertSingle$ is \code{\small assertTrue(widthNoBold <= widthBold)}. 
\end{exmp}


\subsubsection{Step 3: Codify Constituents of MR}
\begin{figure}[!t]
\centering
\resizebox{\linewidth}{!}{\includegraphics{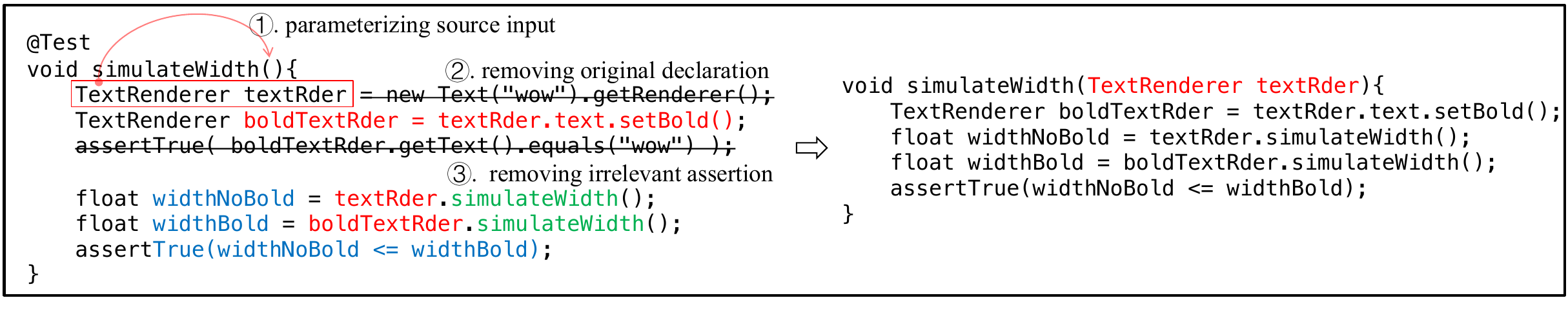}}
\caption{Illustration of constructing a codified MR}\label{fig:codification-example} 
\end{figure}

This step mainly consists of parameterizing the source input and removing irrelevant assertions. We illustrate the process of constructing a codified MR using the example shown in Figure~\ref{fig:codification-example}.

An MTC is in the form of a Java method (because a JUnit test case is formatted as a method). 
To codify MRs as methods parameterized with source inputs, \tool modifies the MTC under codification to take the source input as a parameter. As shown in \Circled{1} of Figure~\ref{fig:codification-example}, the source input \code{\exampleSI} is transformed into a parameter to receive new input values. 
\tool also removes the source input declaration statements (\Circled{2} in Figure~\ref{fig:codification-example}).

Next, \tool removes irrelevant assertions (\Circled{3} in Figure~\ref{fig:codification-example}). 
Assertions not identified as relation assertions are considered irrelevant and removed. 
These irrelevant assertions may be specific to the original value of the source input and could lead to false alarms when new inputs are introduced. 
In Figure~\ref{fig:codification-example}, assertion \code{\exampleIrrlevantAssertion} is removed. 
These modifications enable the codified MR method to receive and validate the relation over values of new source inputs and corresponding outputs. 

A codified MR encompasses steps 2-5 of metamorphic testing, involving constructing the follow-up input, executing the target program on both the source input and follow-up input, and validating the output relation across program executions. As a result, automated test case generation can be achieved only when new source inputs are automatically generated to these codified MRs.


\subsection{Phase~3: MR Filtering}\label{sec:app-phase3}
\congying{[August 10, 2023] This subsection is ready for you to revise.} 

We aim to discover MRs to generate new test cases, where codified MRs can serve as test oracles. 
However, codified MRs that are not applicable to new inputs are ineffective for new test generation~\cite{qiu2020theoretical}. 
Therefore, in this phase, \tool filters out low-quality codified MRs that perform poorly \todo{(e.g., leading to false alarms)} in applying to new test inputs.

\label{sec:validMRandValidInput}
\textbf{Criterion.} 
Following previous Zhang et al.'s work on inferring polynomial MRs~\cite{zhang2014search}, \tool considers \textit{an MR that can apply to at least 95\% of valid inputs as a \textbf{high-quality MR}}. 
Differently, Zhang et al.'s work infer MRs for numeric programs (e.g., $\mathit{sin}$, $\mathit{cos}$, and $\mathit{tan}$). 
\todo{All generated inputs are numerical values and inherently valid, satisfying the input constraints.} 
In our domain, however, we are dealing with object-oriented programs whose inputs are not only primitive types but also developer-defined objects. Randomly generated inputs can be invalid. 
To automatically tell whether an input is valid, we observe that the program under test contains checks for illegal arguments, and thus assume that \textit{a \textbf{valid input} for an MR must be accepted by the input transformation and the methods of the class under test}. 
That means the execution of a valid test input must not trigger an exception from the statements of input transformation and the invoked methods of the class under test until reaching the relation assertion statement of a codified MR. 
Note that our checking (not triggering exception) is less stringent than the actual criterion for determining a valid input that satisfies input constraints.
An invalid input might not trigger an exception due to the lack of developer-written checks for illegal arguments and the absence of exception-throwing mechanisms. 
\todo{
When an invalid input reaches an assertion statement, it may violate the output relation of a codified MR and produce false alarms.
This leads to some high-quality MRs being discarded. 
}



After executing the relation assertion statement, if an \code{AssertionError} occurs, it indicates the valid input has failed, and the codified MR cannot apply to this input. On the other hand, if no alarm is raised, that means this input complies with the codified MR, thereby the codified MR is applicable to this input. 

\textbf{Inputs Generation.} 
Many techniques have been proposed to generate test inputs, such as random~\cite{randoop}, search based~\cite{HarmanJZ15, evosuite}, and symbolic execution based techniques~\cite{CadarS13}. 
Following existing works on test oracle assessment and improvement~\cite{DBLP:conf/issta/JahangirovaCHT16, valerioJTP20}, \tool employs EvoSuite~\cite{evosuite} to generate new inputs for codified MRs.
\todo{Different from Zhang et al.'s work where MRs are for $\mathit{sin}$, $\mathit{cos}$, and $\mathit{tan}$ programs, and 1000 new numeric inputs can be easily generated for each MR, in our domain, the inputs of codified MRs are not only primitive types but also developer-defined objects. For MRs with complex objects as inputs, EvoSuite cannot generate a large amount (e.g., 1000) of valid objects as new inputs. 
So we give the same time budget rather than the same amount of inputs for each codified MR.} 
In line with the configuration of previous works~\cite{Lin2021, SBSTCompetition22}, for each codified MR, \tool runs EvoSuite 10 times with different seeds and gives a time budget of 2 minutes for each run. The detailed configuration of EvoSuite can be found on \tool's website~\cite{mr-extractor}. 

Then, \tool executes these test cases (as illustrated in Figure \ref{fig:procedure} (Phase 3)) and analyzes the execution result (i.e., pass or fail). 
Finally, \tool outputs high-quality codified MRs that can apply to at least 95\% of generated valid inputs. 
\hc{There can be other choices for generating input. For example Randoop and various fuzzers. It would be nice to elaborate more on our choice.} \congying{already added.}

%% file: table/relation-assertion-pattern.tex
\begin{tabularx}{\linewidth}{p{1.8cm}|p{6.3cm}p{8cm}} 
\toprule
\textbf{Pattern}
& \textbf{Assertion APIs in \textsc{JUnit}}
& \makecell[l]{\textbf{Examples} } \\
\midrule
\textbf{BoolAssert}
& {\smaller \texttt{assertTrue}, \texttt{assertFalse}}
& \makecell[l]{
\smaller \texttt{assertTrue(Math.abs(}$\exampleEone$\texttt{)}$>$\texttt{Math.abs(}$\exampleEtwo$)\texttt{);}\\
\smaller \texttt{assertTrue(}$\exampleEone$\texttt{.equalsTo(}-$\exampleEtwo$\texttt{));}}          \\
\midrule
\textbf{CompAssert}
& \makecell[l]{\smaller\texttt{assertSame}, \texttt{assertNotSame}, \texttt{failNotSame},\\
\smaller \texttt{assertEquals}, \texttt{failNotEqual}, \texttt{assertArrayEquals}, \\
\smaller \texttt{assertThat}, \texttt{assertIterableEquals}, \texttt{assertLinesMatch}}
& \makecell[l]{
\smaller \texttt{assertEquals}($\exampleEone, \exampleEtwo$);\\
\smaller \texttt{assertEquals(Math.abs(}$\exampleEtwo$\texttt{), Math.abs(}$\exampleEone$\texttt{));}}           \\
\bottomrule
\end{tabularx}

%% file: src/04-evaluation.tex
\section{Evaluation}\label{sec:evaluation}
\congying{[August 10, 2023] This section is ready for you to revise.}

Our evaluation aims to answer the following research questions: 
\begin{itemize}
\item[\textbf{\numRQSoundness}]\textbf{Precision:} Are MTCs discovered by \tool possessing the derived properties of an MTC? (Section \ref{sec:rq-soundness})
\item[\textbf{\numRQQuality}] \textbf{Quality:} How is the quality of \tool codified MRs \todo{in applying to new inputs for automated test case generation}? (Section \ref{sec:rq-quality})
\item[\textbf{\numRQUsefulness}] \textbf{Usefulness:} How useful are  \tool codified MRs in enhancing test adequacy? (Section \ref{sec:rq-usefulness}) 
\item[\textbf{\numRQUnderstandabiltiy}] \textbf{Comprehensibility:} Are MRs codified by \tool{} comprehensible?  (Section \ref{sec:rq-understandability})
\end{itemize}

\textbf{\numRQSoundness} aims to evaluate the precision of \tool in discovering MTCs, i.e., \todo{whether discovered test cases possess the defined properties of an MTC}. 
To answer \numRQSoundness, we first ran \tool, and \numIdentifiedMTCs{} MTCs from \todo{\numPoJwithIdentifiedMTCs} projects were discovered. 
Then, we manually analyzed 164 sampled MTCs. 
\textbf{\numRQQuality} aims to evaluate the quality of \tool codified MRs, by using a set of new test inputs not present in the \todo{filtering phase of \tool}. 
The results also indicate the effectiveness of the \textit{\phaseThree} phase in the methodology. 
\textbf{\numRQUsefulness} is to evaluate the usefulness of codified MR when integrated with automatically generated inputs. 
Specifically, we analyze whether test suites constructed from codified MRs can enhance test adequacy on top of developer-written and EvoSuite-generated test suites. 
\textbf{\numRQUnderstandabiltiy} aims to assess whether \tool codified MRs are easily comprehensible for developers engaged in tasks like test maintenance or migration. 
For this purpose, we conducted a small-scale \todo{qualitative study} on \todo{52} codified MRs. 

\subsection{\editedT{Dataset Preparation}}

We selected open-source projects from GitHub~\cite{GitHub}. We chose public projects meeting these criteria: (i) labeled as a Java project, (ii) having at least 200 stars, and (iii) created after \todo{01-January-2015}. 
These criteria enable us to analyze high-quality and contemporary Java projects that are more likely to use mature unit testing frameworks like JUnit~\cite{junit5} and TestNG~\cite{testng}. 
The number of stars indicates the popularity and correlates with project quality~\cite{borges2018s}. We considered projects created after \todo{01-January-2015} to exclude old projects that might require obsolete dependencies and frameworks, 
\editedT{
while some popular Java projects that were created before 2015 might be excluded. 
}

By 05-April-2022, {7,395} projects met these criteria and were collected for experiments. \editedT{These projects account for 71.49\% of all popular Java projects that have at least 200 stars and were created both before and after 01-January-2015.}
We cloned the latest version of each selected project at that time and collected tests from these projects. 
We considered methods annotated with ``\code{@Test}'' as tests and files containing tests as test files.  
We excluded \todo{3,327} projects without tests. 
At last, we had \todo{4,068} projects, which contained \todo{1,021,129} Java tests in \todo{545,886} test files. These projects comprised \todo{239,724,897} lines of production code and \todo{80,130,804} lines of test code.

\subsubsection{MTC Discovery} \label{sec:rq-prevalence} \label{sec:MTC-dataset}

\editedT{
We ran \tool on each of the \todo{4,068} projects on a machine with dual Intel\textregistered{} Xeon\texttrademark{} E5-2683 v4 CPUs and 256 GB system memory. 
For the \textit{MTC Discovery} phase, \tool took \todo{18} hours and \todo{58} minutes, with an average analysis time of \todo{16.78} seconds per project. 
Finally, \todo{\numIdentifiedMTCs} MTCs in \todo{\numPoJwithIdentifiedMTCs} (\todo{17.23\%}) projects were discovered. 
On average, each project has \todo{16.19} MTCs. 
}

\editedT{As to \todo{3,367 (82.77\%)} projects where no MTC was discovered, our observations suggest a limited presence of test cases encoded with MRs. 
Typically, test cases in these projects are structured to assert whether the actual output of a method under test aligns with the expected output for a given input. 
Moreover, some projects exhibited inadequate testing, having few test cases. This reduces the chances of discovering MTCs. 
Additionally, \tool is designed to discover MRs of a class. This means that \tool focuses on MRs associated with methods within a single class. MRs over multiple classes or at higher levels are out of the scope of \tool.
}

\begin{figure*}[t]
\centering
\begin{subfigure}{0.45\textwidth}
\centering
\includegraphics[scale=0.25]{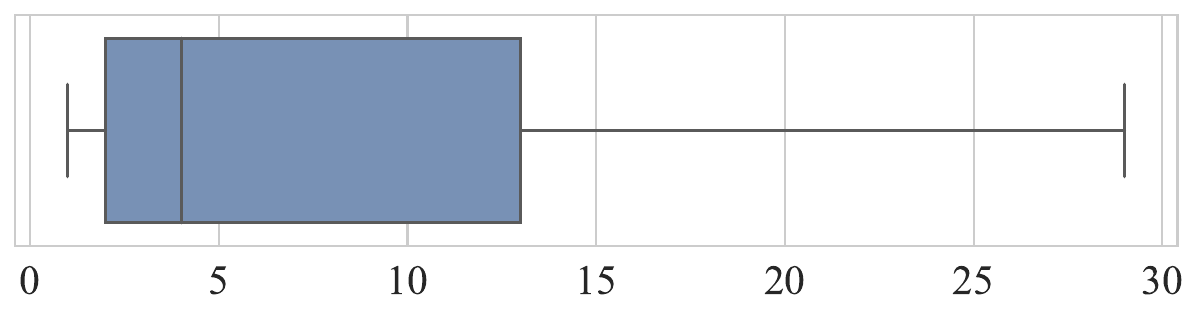}
\caption{Number of MTCs in a project}\label{fig:rq1-prevalence-distribution-number}
\end{subfigure}
\begin{subfigure}{0.45\textwidth}
\centering
\includegraphics[scale=0.28]{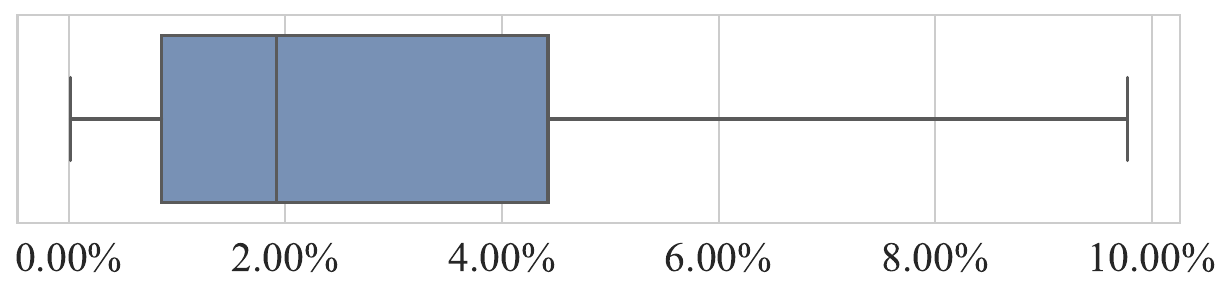}
\caption{Percentage of MTCs in a project}
\end{subfigure}
\caption{Distribution of \todo{\numIdentifiedMTCs} MR-encoded test cases (MTCs) in \todo{\numPoJwithIdentifiedMTCs} projects w.r.t the number and percentage
}\label{fig:rq1-prevalence-distribution-percentage}
\end{figure*}

\textbf{Distribution of MTCs.} The distribution of MTCs provides insights into how MTCs are spread across projects. 
The distribution of \todo{\numIdentifiedMTCs} MTCs in the \todo{\numPoJwithIdentifiedMTCs} projects varies significantly, ranging from a \todo{single} MTC to \todo{500} MTCs.
As shown in Figure~\ref{fig:rq1-prevalence-distribution-number}, the majority of the projects have \todo{1} to \todo{29} MTCs, and the median is \todo{4}. 
Half of the \todo{\numPoJwithIdentifiedMTCs} projects have \todo{2} to \todo{13} MTCs. 
We further analyzed the percentage of MTCs among all tests in each project in Figure~\ref{fig:rq1-prevalence-distribution-percentage}. For the majority of projects, \todo{0.02\%} to \todo{9.78\%} of the tests are MTCs, and the median is \todo{1.91\%}. 
Percentages of MTCs in half of these projects range from \todo{0.8\%} to \todo{4.42\%}. 

We also examined the top 25 projects with the highest number of MTCs (the projects can be found at~\cite{mr-extractor}). These projects span various domains, including complex data structures, data processing, distributed computing, data visualization, smart contracts, website building, code parsing, and more. 
\todo{The results indicate that MTCs are broadly distributed across projects from diverse domains rather than being concentrated within a few projects with specific functionalities.}

\begin{figure}[t]
\centering
\begin{subfigure}{0.35\textwidth}
\centering
\includegraphics[scale=0.26]{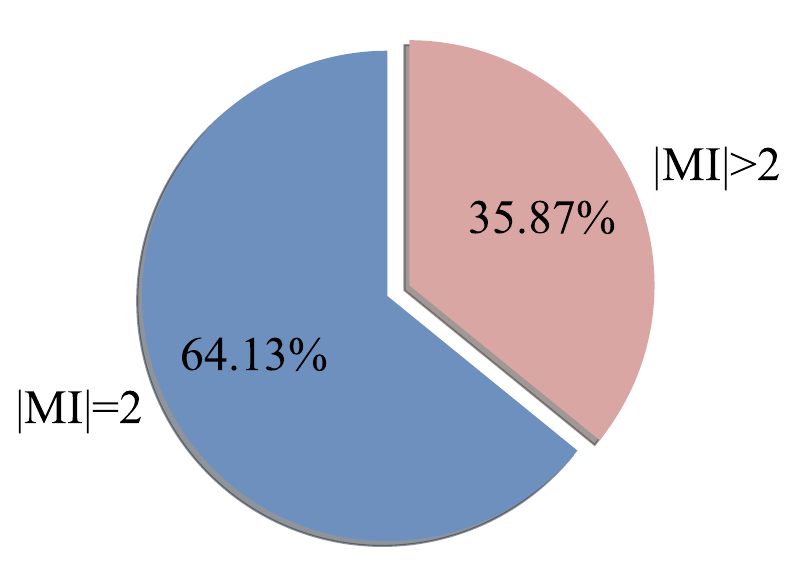}
\caption{Size of involved MI ($|$MI$|$)}\label{fig:rq1-}
\end{subfigure}
\begin{subfigure}{0.35\textwidth}
\centering
\includegraphics[scale=0.26]{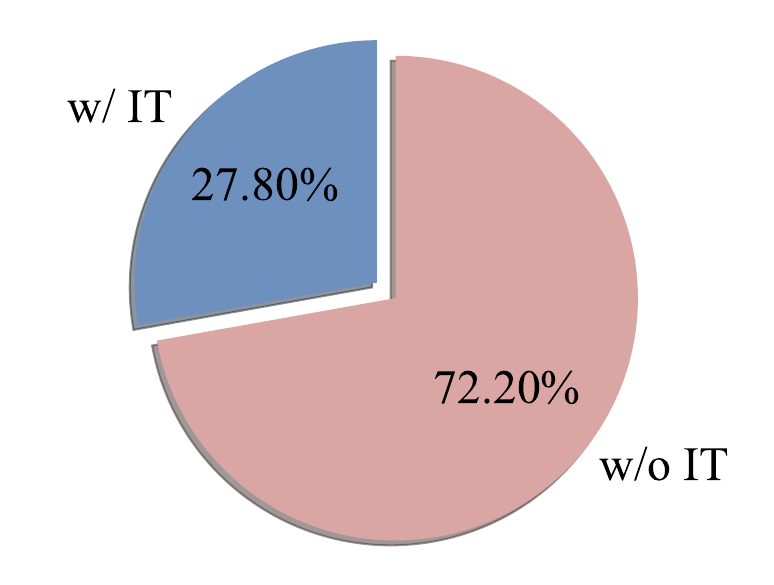}
\caption{Existence of IT, when $|$MI$|$=2}
\end{subfigure}
\caption{\editedTS{Distribution of {21,574} MR instances w.r.t the size of involved method invocations ($|$MI$|$) and the existence of an input transformation (IT)}}\label{fig:rq1-distribution}
\end{figure}

\textbf{Complexity of MTC.}
The discovered \todo{\numIdentifiedMTCs} MTCs contain a total of \todo{21,574} MR instances (introduced in Section \ref{sec:app-phase2}). 
On average, \todo{1.90} MR instances were found per MTC.
\todo{13,836 (64.13\%)} out of the \todo{21,574} MR instances involved only two method invocations.
Among these \todo{13,836} MR instances, \todo{3,847 (27.80\%)} instances in \todo{2,743} MTCs were associated with an input transformation.
This indicates that a significant proportion (\todo{64.13\%}) of MR instances leverage MRs involving only two method invocations, and \todo{72.20\%} of MR instances are without input transformation. 
In this work, we target synthesizing MRs from MR instances involving two method invocations and having input transformation.

These results indicate that numerous MR-encoded test cases are widely spread across open-source projects of different domains. \todo{17.23\%} of projects contain MTCs, and in total \todo{\numIdentifiedMTCs} MTCs were discovered from \numPoJwithIdentifiedMTCs{} projects. 
Besides, the majority of encoded MR instances (\todo{64.13\%}) involve relations with two method invocations. 
The MTC dataset is released and available on \tool's site~\cite{mr-extractor}.


\subsubsection{\editedT{MR Synthesis and Filtering}} \label{sec:dodified-MR-Preparation}
In this paper, we focus on MTCs where MR instances (i) involve two method invocations ($\mathit{|MI|=2}$), (ii) have input transformation, and (iii) are from compilable projects where mutation analysis and EvoSuite-based MR filtering can be conducted. Finally, in \tool discovered MTCs, we collected \todo{485} MTCs from \todo{104} projects. 

In \textit{\phaseTwo} phase, {441} ({90.92\%}) MTCs' encoded MRs were successfully synthesized into compilable codified MRs. 
The other {9.08\%} of MTCs failed due to too complicated external dependencies or code structures. 

In \textit{\phaseThree} phase, \tool filtered candidate codified MRs with inputs generated by EvoSuite.
EvoSuite could successfully generate valid inputs for {125} candidate codified MRs. 
Among {125} candidate codified MRs that have valid inputs, {60.00\%} ({75/125}) passed the \textit{\phaseThree} phase and were finally outputted by \tool as high-quality codified MRs.
The main reasons why some codified MRs were not generated with valid inputs include too complex preconditions, incompatible environment, violation of input constraint, etc., which are detailly discussed in Section~\ref{sec:discuss}.

\subsection{\numRQSoundness: Precision}\label{sec:rq-soundness}
\subsubsection{Experiment Setup}
\tool{} discovers MR-encoded test cases based on static analysis of the source code. 
However, factors such as aliasing, path sensitivity, dynamic language features like reflection, and handling of recursions can cause imprecise analysis results.
Therefore, in \numRQSoundness{}, we aim to evaluate whether \tool{} is precise in discovering MR-encoded test cases in real-world projects. The results also reflect the reliability of our released dataset of discovered MTCs.

To achieve this goal, we manually validate if the MTCs discovered by \tool{} have the two properties mentioned in Section~\ref{sec:criteria-MTC}. 
However, there are \numIdentifiedMTCs{} MTCs discovered in our evaluation subjects, and it is infeasible to check all of them manually. 
\editedT{
Therefore, we randomly sample 164 out of 11,350 MTCs to estimate the precision of \tool{}.
Such a sample size can be calculated by an online calculator~\footnote{https://www.qualtrics.com/experience-management/research/determine-sample-size/}, ensuring a confidence level of 99\% and a confidence interval of 10\% for our estimation result~\cite{heckert2002handbook}.
}

During the validation, two authors of this paper independently inspected the source code of the discovered test cases.
Based on their understanding, they labeled a test case with one of the following labels:
\begin{itemize}
    \item \emph{True Positive} indicates a test case possesses the two properties mentioned in Section~\ref{sec:criteria-MTC}.
    \item \emph{False Positive} indicates a test case does not possess the two properties.
    \item \emph{Unclear} indicates the author cannot understand a test case or tell if the two properties are possessed.
\end{itemize}
After independent labeling, the two authors discussed test cases labeled differently or labeled as \emph{Unclear} and finally reached a consensus.

\subsubsection{Result}
During the manual validation, there were 27 test cases assigned with different labels by the two authors, and the divergences were resolved. 
Finally, among the 164 sampled test cases, 160 cases were labeled as true positives, whereas the remaining four were labeled as false positives. 
Overall, based on a sampled dataset, \tool{} demonstrates an estimated precision of 97\% (with a confidence level of 99\% and a margin of error at 10\%) in discovering MTCs.    

\begin{figure}[!t]
\centering
\resizebox{0.8\linewidth}{!}{\includegraphics{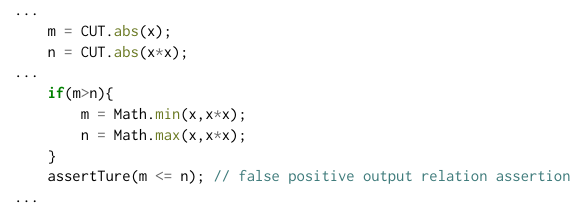}}
\caption{Simplified example of a false positive MTC}
\label{lst:FP-mtc}
\end{figure}

All of the four false positives were due to incorrect identification of $\mathit{\propertyTwo}$.
This is because \tool does not well handle the scopes of variables in complicated cases with re-assigned variables and non-sequential control flows. 
Figure~\ref{lst:FP-mtc} shows a simplified example, where \code{m} and \code{n} are assigned with the return values of method \code{CUT.abs} in the class under test. 
When encountering assertion \code{AssertEquals(m,n)}, \tool mistakenly considers this assertion to fulfill \propertyTwo\ --- validating the relation over outputs of \code{CUT.abs(x)} and \code{CUT.abs(x*x)}. 
Consequently, \tool falsely considers this test case to be positive. 
However, before assertion, the variables \code{m} and \code{n} may be re-assigned with the return values of \code{min(x,x*x)} and \code{max(x,x*x)} which are not methods in the class under test. \code{AssertEquals(m,n)} is a false positive output relation assertion. 


Despite a minor fraction of false positives, most discovered MTCs satisfy the properties mentioned in Section~\ref{sec:criteria-MTC}.
The results show that \tool{} can effectively discover MTCs in real-world projects, and our dataset is of high reliability. 
\begin{answertorq}
\tool{} is precise in discovering MTCs in real-world projects, achieving \editedT{an estimated} precision of 97\% in discovering MTCs. Such high precision indicates the high reliability of our released dataset of MTCs.
\end{answertorq}

\subsection{\numRQQuality: Quality}\label{sec:rq-quality}
\subsubsection{Experiment Setup}
In \textit{\phaseThree} phase, \tool filters out low-quality MRs with new inputs generated by EvoSuite within a certain time budget. 
This approach differs from Zhang et al.'s work~\cite{zhang2014search}, which infers MRs for numeric programs MRs and generates 1,000 numeric inputs for each MR. 
The reason for our choice is the difficulty of generating a vast quantity of complex objects as inputs for some MRs in our domain. This choice potentially weakens the \textit{\phaseThree} phase in the methodology.
Therefore, in this RQ, we aim to evaluate the quality of \tool codified MRs \todo{in applying to new inputs for automated test case generation}. The result also indicates the effectiveness of the \textit{\phaseThree} phase.



To achieve this goal, we reuse the criterion of \textbf{a high-quality MR} defined in Section~\ref{sec:app-phase3}. 
MRs that are not of high quality are termed low-quality MRs. Besides, considering the \textit{\phaseThree} phase has already relied on EvoSuite-generated inputs, we re-ran EvoSuite with different seeds and had a replication check to construct a set of different test inputs for evaluation, thereby mitigating the circularity issue in the evaluation.

\begin{figure}[t]
\centering
\resizebox{0.6\linewidth}{!}{\includegraphics{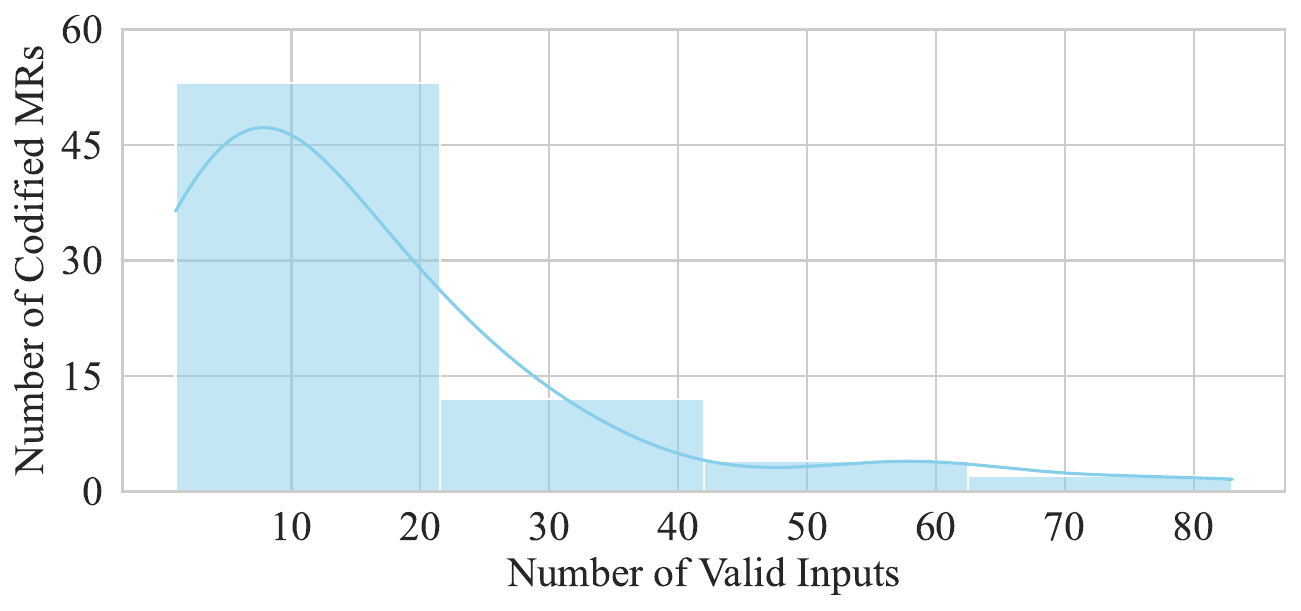}}
\caption{Distribution of generated valid inputs}\label{fig:rq-quality-distribution-valid-inputs}
\end{figure}

In this RQ, we evaluate the quality of \todo{75} codified MRs that are output by \tool after filtering (Section~\ref{sec:dodified-MR-Preparation}). 
In line with the configuration of the previous studies~\cite{Lin2021, SBSTCompetition22}, we re-ran EvoSuite 10 times with different seeds to mitigate the randomness issue on the evaluation results and gave a time budget of 2 minutes for each run\footnote{The detailed configuration of EvoSuite can be found on \tool's website~\cite{mr-extractor}.}. 
With the generated inputs, we filtered out inputs that appeared in the \textit{\phaseThree} phase of \tool, and filtered out invalid inputs according to the criterion of a valid input in Section \ref{sec:validMRandValidInput}. 
Finally, \todo{4} codified MRs did not have newly generated valid inputs. 
\todo{71} codified MRs had \todo{1,995} generated inputs, where \todo{57.69\% (1,151)} of them are valid inputs, with an average of \todo{16.21} valid inputs for each codified MR. 
Figure~\ref{fig:rq-quality-distribution-valid-inputs} shows the distribution of generated test inputs. 


\subsubsection{Result} 
Out of \todo{71} \tool output MRs, we found that \todo{97.18\% (69)} of MRs are high-quality and even applicable to all valid inputs. 
\todo{Two} codified MRs are low-quality. 16 (out of 24) valid inputs of the two codified MRs result in \code{AssertionError} alarms. 
After manually analyzing, we found that the \todo{2} codified MRs are indeed of low quality. For example, the simplified MR $\mathit{width(text) < width(text.setBold())}$ asserts that the layout of bold text should be wider than non-bold text. However, this MR cannot apply when a text is empty or contains only characters that cannot be bold (e.g., ``$\mathit{<>}$'') or the original text is already bold.

\begin{answertorq}
The \textit{\phaseThree} phase in our methodology is effective. 
\todo{97.18\%} of \tool synthesized MRs are of high quality and applicability to new inputs for automated test case generation, demonstrating the practical applicability of \tool.

\end{answertorq}



\subsection{\numRQUsefulness: Usefulness}\label{sec:rq-usefulness}
\subsubsection{Experiment Setup}
In this RQ, we aim to evaluate the practical application of MRs synthesized by \tool when combined with automatically generated test inputs. 
Specifically, one application scenario of synthesized MRs is testing the original programs where these MRs are found, and we focus on assessing how useful codified-MRs-based tests are in complementing existing tests and improving the test adequacy of these original programs. 
    

\textbf{Metrics and Baselines.} We employ the following four metrics to measure the test adequacy. 
\begin{itemize}
\item Line Coverage: the percentage of target programs' lines executed by a test suite.
\item Test Strength: the percentage of executed mutants killed by a test suite.
\item Percentage (\%) of Covered Mutants: the percentage of mutants executed by a test suite.
\item Mutation Score: the percentage of mutants killed by a test suite.
\end{itemize} 


Firstly, codified MRs are integrated with automatically generated valid inputs in the {\numRQQuality} (Quality) to construct codified-MR-based test suites (denoted as \codifiedMRBasedTestSuite). Then, we compare the performance of the codified-MR-based test suite on these four metrics against two baselines: (i) developer-written test suites (\developerWrittenTestSuite) and (ii) EvoSuite-generated test suites (\evosuiteGeneratedTestSuite). 
\editedA{
Note that both the developer-written test suite and the EvoSuite-generated test suite target all methods in the class under test, while a codified MR only invokes MR-involved methods, which is a subset of all methods in the class under test.}
Thus, we do not directly compare the performance of the codified-MR-based test suite against developer-written or EvoSuite-generated test suites. Instead, we investigate whether the codified-MR-based test suites can enhance the test adequacy on top of developer-written and EvoSuite-generated test suites.

We successfully ran PIT~\cite{pitest}, a mutation testing tool, to generate \todo{2,170} mutants for \todo{51} target classes of \todo{75} codified MRs (which were collected in the dataset preparation, Section~\ref{sec:dodified-MR-Preparation}). There are a total of 4,701 lines of code in these target classes.

\editedT{
\textbf{Statistical Analysis.}
We performed a statistical analysis (i.e., Mann-Whitney U-test~\cite{rice2006mathematical, DBLP:journals/stvr/ArcuriB14}) to test the hypothesis --- the fault detection capability of test suites augmented with codified MR-based tests (i.e., \cde{}) is better than existing tests (i.e., \de{}).
Specifically, we compare the fault detection capability based on killed mutants. 
For each mutant, if it is killed by a test suite, the score for this mutant is 1, otherwise 0. Finally, for the Mann-Whitney U-test, test suites \cde{} and \de{} will get a list of scores for all 2170 mutants, respectively. 
\\
}

\subsubsection{Result}

\begin{table*}[t]
\footnotesize
    \caption{Enhancement of test adequacy by codified-MR-based test suites (\codifiedMRBasedTestSuite) on top of developer-written (\developerWrittenTestSuite) and EvoSuite-generated test suites (\evosuiteGeneratedTestSuite)}\label{tab:rq-usefulness}
    \centering
    \input{table/rq-usefulness}

\end{table*}

\begin{figure*}[t]
\centering
\begin{subfigure}{0.24\textwidth}
\centering
\setlength{\leftskip}{-12pt}
\includegraphics[scale=0.43]{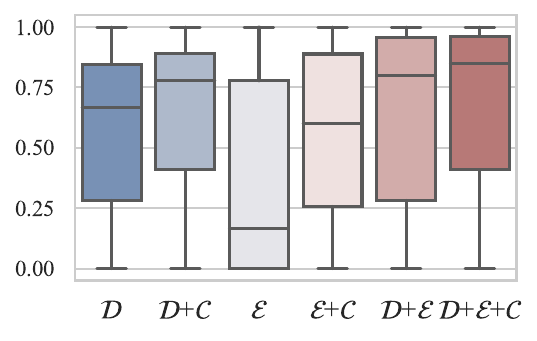}
\caption{Line Coverage}
\end{subfigure}
\begin{subfigure}{0.23\textwidth}
\centering
\includegraphics[scale=0.43]{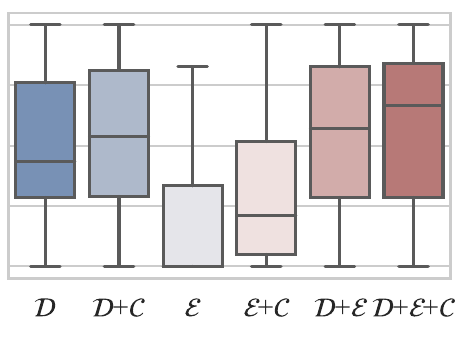}
\caption{Mutation Score}
\end{subfigure}
\begin{subfigure}{0.23\textwidth}
\centering
\setlength{\leftskip}{5pt}
\includegraphics[scale=0.43]{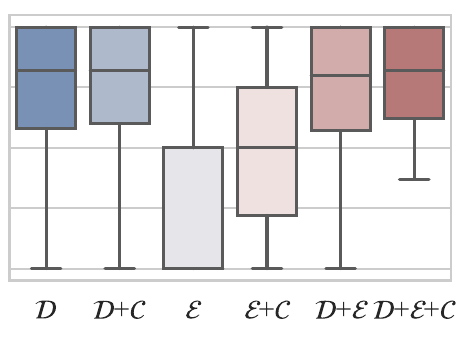}
\caption{Test Strength}
\end{subfigure}
\begin{subfigure}{0.24\textwidth}
\centering
\setlength{\leftskip}{10pt}
\includegraphics[scale=0.43]{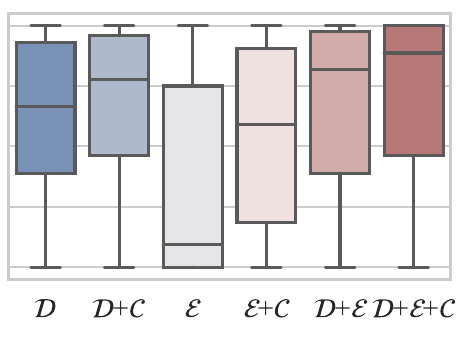}
\caption{\% of Covered Mutants}
\end{subfigure}

\caption{Enhancement of test adequacy by codified-MR-based test suites (\codifiedMRBasedTestSuite) on top of developer-written (\developerWrittenTestSuite) and EvoSuite-generated test suites (\evosuiteGeneratedTestSuite)}\label{fig:rq1-usefulness-boxplot}
\end{figure*}

\begin{figure}[t]
\centering
\begin{subfigure}{0.33\textwidth}
\centering
\includegraphics[scale=0.26]{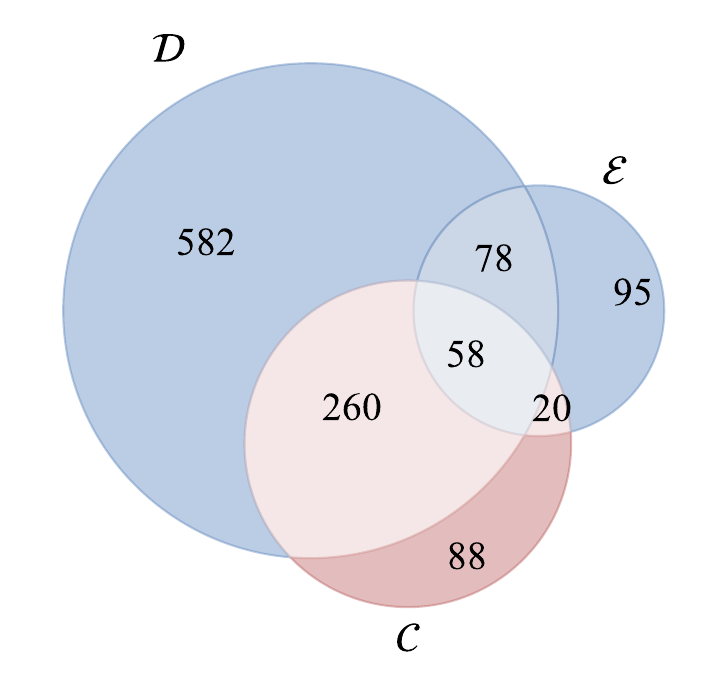}
 \caption{Killed mutants}\label{fig:rq1-usefulness-venn-killed-mutants}
\end{subfigure}
\begin{subfigure}{0.33\textwidth}
\centering
\includegraphics[scale=0.26]{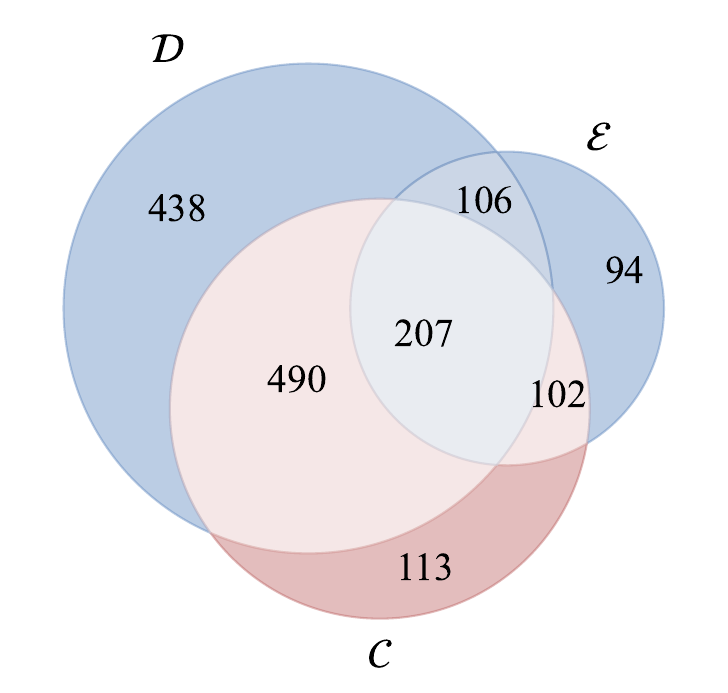}
\caption{Covered mutants}\label{fig:rq1-usefulness-venn-covered-mutants} 
\end{subfigure}

\caption{Comparison of covered and killed mutants by developer-written (\developerWrittenTestSuite), EvoSuite-generated (\evosuiteGeneratedTestSuite), and codified-MR-based (\codifiedMRBasedTestSuite) test suites}\label{fig:rq1-usefulness-venn}
\end{figure}

Table \ref{tab:rq-usefulness} presents the results of the four metrics on \todo{51} classes. 
Compared with \developerWrittenTestSuite, incorporating \codifiedMRBasedTestSuite{} leads to a \todo{13.52\%} increase in the line coverage, and \todo{13.37\%} and \todo{9.42\%} increases in the percentage of covered mutants and mutation score. Compared with \evosuiteGeneratedTestSuite, incorporating \codifiedMRBasedTestSuite{} leads to a remarkable \todo{82.8\%} increase in mutation score and \todo{52.10\%} in line coverage. Even compared with the test suites combining \developerWrittenTestSuite{} and \evosuiteGeneratedTestSuite{} (i.e., \developerWrittenTestSuite+\evosuiteGeneratedTestSuite), incorporating \codifiedMRBasedTestSuite{} can still achieve \todo{6.83\%} and \todo{7.93\%} enhancement in line coverage and mutation score. 
The result indicates that test suites constructed from \tool discovered MRs can effectively improve the line coverage and mutation score, showing the fault-revealing capability of test suites.

Figure~\ref{fig:rq1-usefulness-boxplot} presents box-and-whisker plots showing the comparison results of test suites (\codifiedMRBasedTestSuite, \developerWrittenTestSuite{}, and \evosuiteGeneratedTestSuite{}) on the four metrics. 
We can find that no matter compared with \developerWrittenTestSuite{} or \evosuiteGeneratedTestSuite{} or \developerWrittenTestSuite+\evosuiteGeneratedTestSuite{} test suites, incorporating \codifiedMRBasedTestSuite{} leads to an overall enhancement in terms of the median, first quartile, third quartile, upper and lower whiskers (1.5 times IQR) of four metrics.

\editedT{
Figure~\ref{fig:rq1-usefulness-venn} illustrates the numbers of mutants covered and killed by developer-written (\developerWrittenTestSuite), EvoSuite-generated (\evosuiteGeneratedTestSuite), and codified-MR-based (\codifiedMRBasedTestSuite) test suites. 
Codified-MR-based test suites cover 215 (+11.37\%) more mutants and kill 108 (+9.42\%) more mutants, compared with existing developer-written test suites.  
Even compared with the combination of developer-written and EvoSuite-generated test suites, codified-MR-based test suites have 113 (+6.95\%) exclusively covered mutants and 88 (+7.93\%) exclusively killed mutants.
Furthermore, the result of the $t$-test shows that our hypothesis is retained. The fault detection capability of test suites augmented with codified MR-based tests (i.e., \cde{}) is significantly better than existing tests (i.e., \de{}) ($p$-value=0.003 < 0.05 which is a typical threshold of significance). 
The corresponding effect size (i.e., normalized U statistic) is 0.52. 
These results indicate the usefulness of codified MRs in enhancing the test adequacy (i.e., the test coverage and fault-detection capability). 
} 

The enhanced test adequacy by codified-MR-based test suites results from the effective integration of high-quality test oracles (i.e., codified MRs) with a set of diverse test inputs. 
In developer-written test suites, although test oracles are well-crafted and invaluable, each oracle typically applies to one test input. 
EvoSuite-generated test suites have a large number of test inputs but fall short in the quality of test oracles~\cite{evosuite2011, Fraser2013EvosutieTSE}.

Codified-MR-based tests merge the merits of both developer-written tests and EvoSuite-generated tests. 
When compared with developer-written tests, codified-MR-based tests leverage the same reliable test oracles but with a greater diversity of random test inputs that explore more branches of the target programs. 
When compared with EvoSuite-generated tests, codified-MR-based tests do not have a higher quantity of test inputs, but offer rich developer-crafted test oracles and more meaningful sequences of method invocations (since codified MRs are structured by at least two method invocations, i.e., \propertyOne{} in Section~\ref{sec:necessary properties of MT}). 
EvoSuite was designed to generate only five types of assertions~\cite{evosuite2011}.
Nevertheless, EvoSuite's random generation of test inputs and sequences cannot succeed in invoking some methods that require complex pre-conditions. 
The corresponding examples and detailed analysis can be found in \tool's website~\cite{mr-extractor}. 
As a result, codified-MR-based test suites can effectively improve both line coverage and mutation score.

\hc{The result discussion is too short. We may discuss our implications given the evaluation results. This also applies to other RQs.}
\congying{it is better now? or do you have any suggested ways to discuss the implications?}
\hc{It would be nice if we can pick an example to discuss how our technique help automated MT and thus enhancing the quality of the test suite.
Also, we can discuss a few examples where our technique fail to do so, and investigate the reason behind.
A good evaluation should show both the strength and weakness of the technique.
This also applies to other RQs.}


\begin{answertorq}
    \editedT{
    Test cases constructed from codified MRs lead to \code{13.52\%} and \code{9.42\%} increases in line coverage and mutation score for programs with developer-written test suites, demonstrating the practical usefulness of codified MR in complementing existing tests and enhancing test adequacy. 
    }
\end{answertorq}

\subsection{\numRQUnderstandabiltiy: Comprehensibility}\label{sec:rq-understandability}
\subsubsection{Experiment Setup}
\todo{We consider that \tool synthesized MRs are useful not only for testing their original programs but also for testing other programs that share similar functionalities.} 
In such usage scenarios, when an MR is easy to understand, it simplifies the debugging and maintaining processes. 
Furthermore, comprehensible MRs facilitate test migration for other programs with similar functionalities.
Therefore, we design \numRQUnderstandabiltiy{} to assess the comprehensibility of codified MRs synthesized by \tool.

To this end, we conducted a small-scale qualitative study with five PhD participants who are experienced in programming in Java and MT. 
Specifically, all participants have more than one year of experience researching MT-related topics, and more than three years of programming in Java and using JUnit. 



\textbf{Procedure.} 
To reduce manual efforts,
we randomly sampled \todo{52} cases from the \todo{75} \tool synthesized codified MRs (which were collected in the dataset preparation, Section~\ref{sec:dodified-MR-Preparation}) for the qualitative study. 
Such a sample size can be calculated by an online calculator~\footnote{https://www.qualtrics.com/experience-management/research/determine-sample-size/}, ensuring a confidence level of 99\% and a confidence interval of 10\% for our analysis result~\cite{heckert2002handbook}.

For each codified MR, the participants were required to understand (i) the logic of the MR and (ii) the relevance of this MR to the class under test. Then, the participants rate the comprehensibility of this MR. 
To avoid neutral answers, participants express their opinions using a 4-point Likert scale~\cite{Pedro2023} (i.e., 1: very difficult to understand, 2: difficult to understand, 3: easy to understand, and 4: very easy to understand). 


\editedT{
\textbf{Statistical Analysis.}
After participants rated the comprehensibility of sampled MRs, we performed a statistical analysis (i.e., one-sample $t$-test~\cite{rice2006mathematical, DBLP:journals/stvr/ArcuriB14}) on the rating results. 
The one-sample $t$-test is a statistical method to test hypotheses about whether the mean of one group of samples differs from a given value. 
}

\editedT{
We first aggregated the ratings for each codified MR.
Specifically, for each codified MR, we calculated the average of the comprehensibility scores given by the raters (denoted as $\bar{X}$).
Then, we tested the hypothesis --- the mean of \(\bar{X}\){} over the sampled MRs is greater than 2.5, where 2.5 represents a neutral score.
}

\subsubsection{Result}~\label{sec:comprehensibility-result} 
\begin{figure}[t]
\centering
\resizebox{0.68\linewidth}{!}{\includegraphics{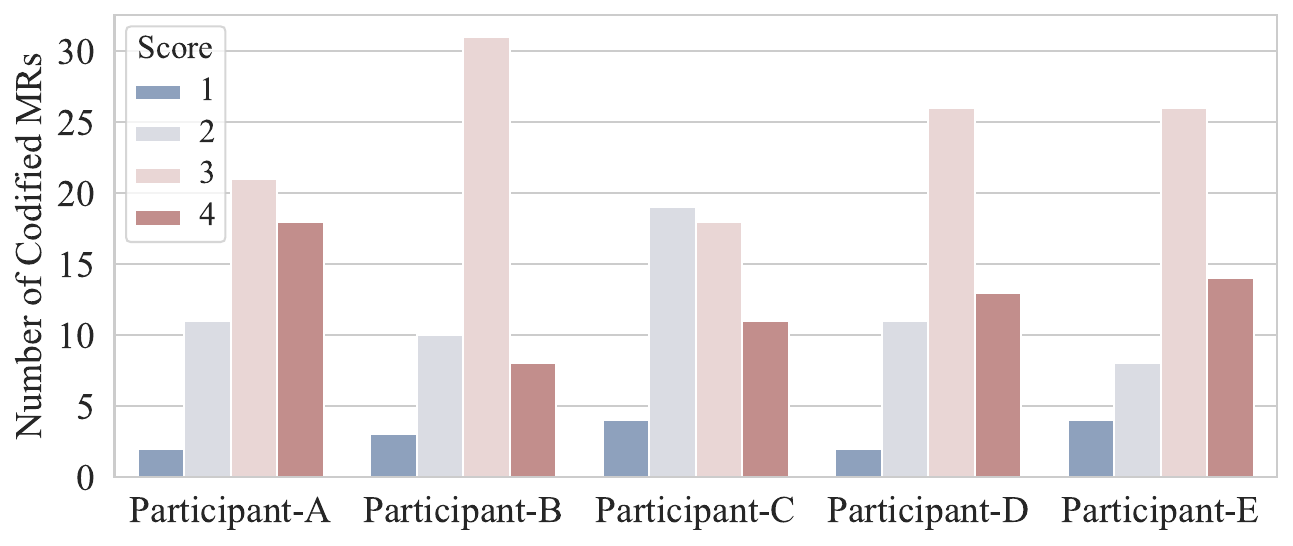}}
\caption{Comprehensibiliy scores of \todo{52} \tool synthesized MRs (Score: 1. very difficult, 2. difficult, 3. easy 4. every easy to understand)}\label{fig:rq-understandability}
\end{figure}

Figure~\ref{fig:rq-understandability} shows the participants' responses to the comprehensibility of codified MRs. Overall, \todo{55.76\% to 76.92\%} of the sampled codified MRs are easy (or very easy) for participants to understand. Moreover, \todo{15.38\% to 34.61\%} of codified MRs are scored as very easy. 
However, there are still \todo{23.08\% to 44.24\%} of the sampled codified MRs that are difficult (or very difficult) to understand. 
\editedT{
The result of the one-sample $t$-test shows that our hypothesis is retained. 
Specifically, the mean comprehensibility of sampled MRs is significantly greater than the neutral score $\mu$=2.5 ($p$-value=3.46$\times 10^{-6}$ $<$ 0.05 which is a typical threshold of significance), and the corresponding effect size (i.e., Cohen's $d$) is 0.70. 
These results indicate that codified MRs are comprehensible. 
}

We also gathered feedback from participants to investigate the factors that make synthesized MRs difficult to understand. 
We found that the main difficulty in understanding some MRs is from the complexity of certain classes under test. 
The test cases in our evaluation were collected from highly-starred Java projects, which often exhibit complex structural dependencies between classes (Section~\ref{sec:MTC-dataset}). 
In our qualitative study, participants were required to understand the relevance between an encoded MR and the class under test. 
Some classes are too complicated for participants to understand their functionalities and business logic, thus making it difficult to understand the relevance. 
However, it is important to note that, for developers who actively maintain these projects or seek to migrate these test oracles (i.e., codified MRs) to similar functionalities in other programs, the codified MRs might be relatively simpler to understand. Familiarity with the projects would likely mitigate the difficulties posed by class complexity. 



\begin{answertorq}
\todo{55.76\% to 76.92\%} of codified MRs can be easily comprehended, showcasing the potential of codified MRs for practical adoption by developers engaged in test maintenance and migration.
\end{answertorq}

\section{Discussion} \label{sec:discuss}
\subsection{Threats to Validity}
We have identified potential threats to the validity of our experiments and have taken measures to mitigate them.

    \textbf{Subjectivity in Human Judgment.}
The evaluation of precision (\numRQSoundness) and comprehensibility (\numRQUnderstandabiltiy) depends on human judgment. To reduce potential subjectivity and misjudgments, we gave the participants a training session before manual validation. for \numRQSoundness, two authors independently validated samples, and then collaboratively resolved any uncertainties or disagreements and came to a consensus, ensuring a rigorous cross-checking mechanism. 
For \numRQUnderstandabiltiy, participants without sufficient experience in MT, Java, and JUnit may affect the results. To mitigate the threats, all involved participants had a solid background in MT, Java, and JUnit, establishing a consistent level of expertise as a baseline for evaluation. 

    \textbf{Sampling Bias.}
The evaluation of precision (\numRQSoundness) and comprehensibility (\numRQUnderstandabiltiy) is based on randomly sampled cases. Different samples may result in different results. 
To mitigate this threat, our sample size statistically ensures a confidence level of 99\% and a confidence interval of 10\% for our evaluation result.

    \textbf{Representativeness of Experiment Subjects.}
A possible threat is whether our findings on the selected OSS projects can be generalized to other popular projects. 
To mitigate this threat, we first adopted criteria from previous empirical studies on OSS projects~\cite{DBLP:conf/icsm/Wang0HSX0WL20,DBLP:journals/ese/HuangCXWSPWL22} to select high-quality and well-maintained Java projects, as described in Section~\ref{sec:rq-prevalence}. 
\editedT{
Then, we quantified the coverage of our selected projects to all popular Java projects on GitHub. 
The result shows that our selected projects account for 71.49\% of all popular projects that have at least 200 stars and were created before the cut-off date of our evaluation (i.e., 05-April-2022).
This coverage suggests that our selected projects are representative.
}


    \textbf{EvoSuite Configuration.}
The choice of parameters for EvoSuite, such as search budget, time limit, and seeds, might affect valid inputs generated by EvoSuite. 
When EvoSuite generates different valid inputs for evaluation, the results of the quality (\numRQQuality) and usefulness (\numRQUsefulness) of codified MRs can be different. To mitigate this threat, we followed the practices of existing studies~\cite{Lin2021, SBSTCompetition22} to run EvoSuite 10 times and chose appropriate parameters that fit our scenario.

\subsection{\editedT{Applications of \tool}}
Metamorphic testing is an approach to both test result verification (i.e., test oracle problem) and test case generation~\cite{chen2018metamorphic} based on metamorphic relations (MRs). 
\tool aims to synthesize MRs from existing test cases that encode domain knowledge and suggest useful MRs. 
Such MRs are useful for testing not only their original programs but also other programs that share similar functionalities. 

As to testing original programs, \tool synthesized MRs help test case generation. 
Synthesized MRs are in the form of parameterized methods, which can be easily integrated with automated input generators to enable automated test case generation. 
This results in a higher fault-detection capability (as evaluated in Section~\ref{sec:rq-usefulness}: Usefulness). 
Furthermore, codified MRs, representing properties of target programs, help describe the behaviors of classes under test across potential test inputs, simplifying test maintenance.



\subsection{Limitations and Future Work} 
Despite \tool being effective in discovering and synthesizing high-quality and useful MRs from existing test cases, \tool still has several limitations. 
\begin{enumerate}
    \item \tool only considers MR instances that involve exactly two method invocations because the constituents of their encoded MRs \editedT{are unambiguous and easier to identify} than MR instances involving more than two method invocations. 
    Synthesizing MRs from instances that involve more than two method invocations can be challenging and interesting future work.
    \item \tool only considers MTCs with explicit input relation (i.e., input transformations). 
    Synthesizing MRs from MTCs without explicit input relations could be challenging and interesting future work. 

    \item \tool statically analyzes the source code of test cases. Factors such as aliasing, path sensitivity, and dynamic language features can cause imprecise analysis results, as discussed in Section~\ref{sec:rq-soundness}. Our sampling result (4 false positives out of 164 samples) reveals that this problem is relatively minor in practice. 
    
    \item 
    \tool synthesized MRs can be of low quality and cause false alarms. To discard such low-quality MRs, we designed a filtering phase based on the pass ratio (i.e., at least 95\%) of valid inputs. 
    However, \tool determines the validity of an input based on developer-written checks (such as \code{IllegalArgumentException} statements).
    When such checks are lacking, invalid test inputs may reach assertion statements, violate the output relation, and produce false alarms. Due to the lack of checks for invalid test inputs, \tool cannot differentiate between false alarms and true bug-exposing alarms. The filtering phase in \tool may discard some high-quality and bug-exposing MRs. 
    Effectively distinguishing the validity of an input and assessing the quality of MRs could be interesting future work. 
    
    \item \tool employs EvoSuite-generated inputs to evaluate the quality and usefulness of codified MRs. However, EvoSuite is coverage-based and ineffective in generating a large number of valid inputs. Here are several main reasons~\cite{Fraser2013Evosuite}: 
(i) Complex precondition of codified MRs: the time budget may not be enough for EvoSuite to construct complex objects that involve many dependencies or deep hierarchies; 
(ii) Incompatible environment: EvoSuite can be incompatible with some libraries or dependencies in the target project; 
(iii) Bugs of EvoSuite: EvoSuite has bugs that cause crashes during generating inputs for codified MRs; 
(iv) Violation of input constraint: some EvoSuite-generated inputs did not conform to the expected input format or constraints (e.g., strings meeting the ``mm/dd/yy'' date format); 
(v) Invalid call sequence: the precondition for invoking a method is not satisfied (e.g., the requirement of invoking \code{setup()} first is not satisfied in the EvoSuite-generated test sequence). 
As noted in~\cite{Fraser2013Evosuite}, ``other prototypes are likely to suffer from the same problems we face with EvoSuite.'' Generating complex objects in the real world remains a challenge for automatic tools. 

\end{enumerate}




%% file: table/rq-usefulness.tex

\setlength{\tabcolsep}{2.1pt}
\begin{tabularx}{\linewidth}{l|ccc|ccc|ccc}
\toprule
\multirow{3}{*}{\textbf{Metrics}} &
\multicolumn{3}{c}{\textbf{VS. \developerWrittenTestSuite}} &
\multicolumn{3}{c}{\textbf{VS. \evosuiteGeneratedTestSuite}} &
\multicolumn{3}{c}{\textbf{VS. \developerWrittenTestSuite+\evosuiteGeneratedTestSuite}} \\
\cmidrule(l{1pt}r{1pt}){2-4} 
\cmidrule(l{1pt}r{1pt}){5-7} 
\cmidrule(l{1pt}r{1pt}){8-10} 
& \textbf{\developerWrittenTestSuite} & \textbf{\developerWrittenTestSuite+\codifiedMRBasedTestSuite} & \textbf{Enhancement}
& \textbf{\evosuiteGeneratedTestSuite} & \textbf{\evosuiteGeneratedTestSuite+\codifiedMRBasedTestSuite} & \textbf{Enhancement}
& \textbf{\developerWrittenTestSuite+\evosuiteGeneratedTestSuite} & \textbf{\developerWrittenTestSuite+\evosuiteGeneratedTestSuite+\codifiedMRBasedTestSuite} & \textbf{Enhancement} \\ 
\midrule

Line   Coverage         & 0.5769               & 0.6549               & +13.52\%             & 0.3735               & 0.5682               & +52.10\%              & 0.6351               & 0.6785               & +6.83\%              \\ 
Test Strength           & 0.7162               & 0.7366               & +2.86\%              & 0.2420               & 0.4889               & +102.03\%            & 0.6977               & 0.7369               & +5.62\%              \\
\% of Covered   Mutants & 0.5960               & 0.6757               & +13.37\%             & 0.3675               & 0.5389               & +46.63\%             & 0.6598               & 0.7057               & +6.95\%              \\
Mutation Score          & 0.5032               & 0.5506               & +9.42\%              & 0.1789               & 0.3271               & +82.80\%              & 0.5395               & 0.5823               & +7.93\%              \\ 
\bottomrule
\end{tabularx}

%% file: src/05-related.tex

\section{Related Work}
\subsection{MR Identification}
Many studies proposed MRs for testing programs of various domains (e.g., compilers~\cite{donaldson2016metamorphic,Donaldson19,xiao2022metamorphic,ma2023fuzzing}, quantum computing~\cite{paltenghimorphq}, and AI systems~\cite{applis2021assessing,wang2019metamorphic,ma2020metamorphic,lindvall2017metamorphic,DBLP:journals/ese/TianMWLCZ21,DBLP:journals/tosem/CaoLLWCC22}).
We review and discuss the most closely related work in systematically identifying MR.

\textbf{MR Pattern Based Approaches.}
Segura et al.~\cite{segura2018metamorphic} proposed six MR output patterns for Web APIs, and a methodology for users to identify MRs. 
Similarly, Zhou et al.~\cite{zhou2018metamorphic} proposed two MR input patterns for testers to derive concrete metamorphic relations. 
These approaches simplify the manual identification of MRs but have limitations: (i) MR patterns are designed for certain programs (such as RESTful web API), (ii) requiring manual effort to identify concrete MRs, and (iii) MR patterns only cover certain types of relations (e.g., equivalence) are not general to complicated or customized relations.
In contrast, \tool automatically discovers and synthesizes codified MRs without manual effort and is not limited to MRs of certain programs or certain types. 
Chen et al. proposed METRIC~\cite{chen2016metric}, enabling testers to identify MRs from given software specifications using the category-choice framework. METRIC focuses on the information of the input domain. 
Sun et al. proposed METRIC+~\cite{sun2019metric}, an enhanced technique leveraging the output domain information and reducing the search space of complete test frames.
Differently, \tool synthesizes MRs from test cases and does not require the software specification and test frames generated by the category-choice framework. 

\textbf{MR Composition Based Approaches.}
The MR composition techniques were proposed to generate new MRs from existing MRs.
Qiu et al.~\cite{qiu2020theoretical} conducted a theoretical and empirical analysis to identify the characteristics of component MRs making composite MRs have at least the same fault detection capability. They also derive a convenient, but effective guideline for MR composition.
Different from these works, \tool does not require existing MRs and can complement these approaches by providing synthesized MRs for composition-based approaches.

\textbf{Search and Optimization Based Approaches.}
Zhang et al.~\cite{zhang2014search} proposed a search-based approach for automatic inference of equality polynomial MRs. 
By representing these MRs with a set of parameters, they transformed the inference problem into a search for optimal parameter values.
Through dynamic analysis of multiple program executions, they employed particle swarm optimization to effectively solve the search problem.
Building upon this, Zhang et al.~\cite{zhang2019automatic} proposed AutoMR, capable of inferring both equality and inequality MRs. Firstly, they proposed a new general parameterization of arbitrary polynomial MRs. Then, they adopt particle swarm optimization to search for suitable parameters for the MRs.
Finally, with the help of matrix SVD and constraint-solving techniques, they cleanse the MRs by removing the redundancy. These approaches focus on polynomial MRs, while \tool considers MRs as boolean expressions, allowing for greater generalization.

Ayerdi et al.~\cite{ayerdi2021generating} proposed GAssertMRs, a genetic-programming-based approach to generate MRs automatically by minimizing false positives, false negatives, and the size of the generated MRs.
However, MRs generated by GAssertMRs are limited to three pre-defined MR Input Patterns. 
Sun et al.~\cite{sun2016mumt} proposed a semi-automated Data-Mutation-directed approach, $\mu$MT, to generate MRs for numeric programs. $\mu$MT makes use of manually selected data mutation operators to construct input relations, and uses the defined mapping rules associated with each mutation operator to construct output relations. However, MRs generated by $\mu$MT are limited to pre-defined mapping rules. In comparison, \tool has no such constraints, applicable for more than numeric programs. 
\jr{Just have a quick glimpse. Too many details and categories. Summarize them.}

\textbf{Machine Learning Based Approaches.}
Kanewala and Bieman~\cite{kanewala2013using} proposed an ML-based method that begins with generating a control flow graph (CFG) from a function's source code, extracts features from the CFGs, and then builds a predictive model to classify whether a function exhibits a specific metamorphic relation.
Building upon this, Kanewala et al.~\cite{kanewala2016predicting} further identified the most predictive features and developed an efficient method for measuring similarity between programs represented as graphs to explicitly extract features.
Blasi et al.~\cite{blasi2021memo} introduced MeMo, which automatically derives metamorphic equivalence relations from Javadoc, and translates derived MR into oracles. 
Different from Memo, \tool is not limited to equivalence MRs. 
These approaches rely on source code or documentation to discover MRs. \tool complements these approaches by synthesizing MRs from test cases. 

\jr{I also do not quite understand why you partition these related works by their underlying techniques. What properties do those works under the same category share? How are they related to \tool? }
\hc{I agree with Jiarong. It would be more clear to group the work by the problem they are trying to solve. In this case, we can present how the problem is addressed by people in the community gradually. We can also discuss their strength and weakness, as well as making a comparison with our technique.}
\hc{It would be nice if we can reorganize this section.}

\subsection{\editedT{Parameterized Unit Tests}}\label{sec:related-PUT}
Parameterized unit tests (PUTs) are tests that accept parameters. A single PUT can be executed with varying input values. 
    PUTs offer several advantages in software testing.
    PUTs are applied with a range of test inputs that can be automatically (e.g., using EvoSuite~\cite{evosuite2011}) to exercise paths of the methods under test. 
    The high test coverage typically results in a better fault-detection capability compared to conventional unit tests. 
    Unlike conventional unit tests, PUTs can take parameters that can be bound to a set of values, allowing exploration of more program states by a single test, making maintenance easier, and reducing test redundancy.

    \hc{The transition between paragraphs is too stiff.}\congying{I have tried to make the transition smoother. Is it better now?} 
    Several studies have been conducted to generate PUTs. 
    Fraser et al.~\cite{DBLP:conf/issta/FraserZ11} proposed to generate PUTs from scratch using a genetic algorithm to generate method-call sequences and using mutation analysis to construct test oracles. 
    Kampmann et al.~\cite{DBLP:conf/icse/KampmannZ19} assumed the existence of high-quality system tests and proposed to automatically extract parameterized unit tests from system test executions. 
    Thummalapenta et al.~\cite{DBLP:conf/fase/ThummalapentaMXTH11} proposed a methodology (termed \textsc{Test-Generation}) to help developers retrofit conventional unit tests into PUTs.

    \tool differs from these earlier studies by synthesizing the underlying metamorphic relations from existing unit tests. Additional unit tests can be automatically generated based on the synthesized relations. The methodology of \tool is orthogonal to those adopted by these studies, which have different assumptions and application scenarios.
    Furthermore, Thummalapenta et al.~\cite{DBLP:conf/fase/ThummalapentaMXTH11} aimed to study the costs and benefits of converting unit test cases into parameterized unit tests. The work conducted an empirical study and proposed a methodology to help developers manually promote inputs as parameters, define test oracles, add assumptions, and construct mock objects based on existing test cases. 
    In contrast, \tool automatically synthesizes codified MRs from existing test cases.

%% file: src/06-conclusion.tex
\section{Conclusion}
Developers embed domain knowledge in test cases. Such domain knowledge can suggest useful MRs as test oracles, which can be integrated with automatically generated inputs to enable automated test case generation. 
Inspired by the observation, we introduce \tool to automatically discover and synthesize MRs from existing test cases in OSS projects. 
We model the semantics of MRs using a set of properties. \tool first discovers MR-encoded test cases based on these properties, and then synthesizes the encoded MRs by codifying them into parameterized methods to facilitate new test case generation. Finally, \tool filters out low-quality MRs that demonstrate poor quality in their applicability to new inputs for automated test case generation.

\tool discovered over 11,000 MR-encoded test cases from 701 OSS projects. 
Experimental results show that \tool achieves a precision of 0.97 in discovering MTCs. 
\todo{97.18\%} of the MRs codified by \tool from these test cases are of high quality and applicability for automated test case generation, demonstrating the practical applicability of \tool. 
Moreover, test cases constructed from these synthesized MRs can effectively improve the test coverage of the original test suites in the OSS projects and those generated by EvoSuite, demonstrating the practical usefulness of \tool synthesized MRs. 
Our qualitative study shows that \todo{55.76\% to 76.92\%} of the MRs codified by \tool can be easily comprehended, showcasing the potential of synthesized MRs for practical adoption by developers.

%% file: src/10-statement.tex
\section{Data Availability and Statement}
\textbf{Data Availability.}
We make \tool and the experimental data publicly available at \tool's site~\cite{mr-extractor} to facilitate the reproduction of our study and relevant studies of other researchers in the community. 

\textbf{Statement of AI Tool Usage.}
During the paper writing, we used Grammarly~\cite{grammarly} and ChatGPT~\cite{chatGPT} to check grammar.

%% file: src/10-acknowledgement.tex
\begin{acks}
    We would like to thank the anonymous reviewers for their insightful comments and suggestions. 
    This work is supported by the National Science Foundation of China (Grant No. \code{61932021}), the Hong Kong Research Grant Council/General Research Fund (Grant No. \code{16207120}), and the
    Hong Kong Research Grant Council/Research Impact Fund (Grant No. \code{R5034-18}). 
\end{acks}